\documentclass{aa}  
\usepackage{graphicx}
\usepackage{txfonts}
\begin{document}
   \title{A long hard look at the minimum state of PG 2112+059 with XMM-Newton}

   \author{N. Schartel 
          \inst{1} 
          \and 
          P.M. Rodr\'{i}guez-Pascual 
          \inst{1}
          \and 
           M. Santos-Lle\'{o} 
          \inst{1}  
          \and 
          E.~Jim\'{e}nez-Bail\'{o}n 
          \inst{2}
          \and 
          L. Ballo 
          \inst{1,3} 
          \and 
          E.~Piconcelli 
          \inst{4}
       }          

   \offprints{Norbert.Schartel@sciops.esa.int }

   \institute{XMM-Newton Science Operations Centre, ESA, Villafranca del Castillo, Apartado 78, 
              28691 Villanueva de la Ca{\~nada}, Spain \\ \and  
              Instituto de Astronom\'\i{}a, Universidad Nacional Aut\'onoma de M\'exico, 
              Apartado Postal 70-264, 04510-Mexico DF, M\'exico \\ \and  
              Instituto de F\'{\i}sica de Cantabria (CSIC-UC), E-39005 Santander, Spain  \\ \and  
              Osservatorio Astronomico di Roma (INAF), via Frascati 33, 
                00040 Monteporzio Catone, Italy
             }

   \date{Received ??, ???; accepted ??, 20??}

  \abstract{}
{Our observational aim is to perform a long X-ray observation of the 
 quasar PG~2112+059 in its low or minimum state. 
Starting form this very peculiar emission state, we intend to constrain the intrinsic emission mechanism by comparing new and
  old data, corresponding to different source states.  }
{XMM-Newton successfully detected the minimum state of \object{PG~2112+059}
 during a short snapshot observation and performed a long follow-up observation.
The high signal-to-noise spectra are modelled assuming different emission 
 scenarios and compared with archival spectra taken by XMM-Newton and Chandra. }
{The \object{PG~2112+059} X-ray spectra acquired in May 2007 allowed the detection 
 of a weak iron fluorescent line, which is interpreted as being caused by reflection from
 neutral material at some distance from the primary X-ray emitting source. 
The X-ray spectra of  \object{PG~2112+059} taken at five different epochs during  
different flux states can be interpreted within two different scenarios.
The first consists of two layers of ionised material with column densities of  
 $N_H \sim 5 \times 10^{22} cm^{-2}$ and  $N_H \sim 3.5 \times 10^{23} cm^{-2}$, respectively.  
The first layer is moderately ionised and its ionisation levels follow the flux changes, 
 while the other layer is highly ionised and does not show any correlation with the flux of the source.
The spectra can also be interpreted assuming reflection by an ionised accretion 
 disk seen behind a warm absorber.  
The warm absorber ionisation is consistent with being correlated with the flux of the source, 
 which provides an additional degree of self-consistency with the overall reflection-based model.
We explain the spectral variability with light bending according to the models of 
 Miniutti and Fabian  and constrain the black hole spin to be $a/M > 0.86.$ 
Both scenarios also assume that  a distant cold reflector is 
 responsible for the Fe K  $\alpha$ emission line.}
{Light bending provides an attractive explanation of the different states of \object{PG~2112+059} 
 and may also describe  the physical cause of the observed properties of other X-ray weak quasars.
The observations of \object{PG~2112+059} in different states provide valuable constraints, 
 although are unable to break the degeneracy  between complex absorption scenarios and 
 reflection from an ionised disk.}


   \maketitle

\section{Introduction}
\label{In}

X-ray weak or soft X-ray weak quasars are characterised by an X-ray 
 emission lower, by a factor  of 10-30, than expected based 
 on their optical-UV emission (Laor et al. \cite{Laor1997}; 
 Wang et al. \cite{Wang1996}; Elvis \& Fabbiano \cite{Elvis1984}). 
About 10\% of bright, optically selected quasars belong to this subclass 
 (Laor et al. \cite{Laor1997}; Brandt, Laor and Wills  \cite{Brandt2000};
 Gibson et al. \cite{Gibson2008}).

Analysing the Boroson \& Green (\cite{Boroson1992}) sample of quasars, 
 Brandt et al. (\cite{Brandt2000}) found a correlation between soft X-ray 
 weakness and C IV absorption equivalent width (EW). 
This correlation implies that absorption is the primary cause of soft X-ray 
 weakness, which is in general agreement with models connecting orientation 
 and absorption strength. 
However, a uniform screen of material that completely covers both the X-ray 
 and ultraviolet emission sources is in disagreement with the shape 
 of the correlation (Brandt et al. \cite{Brandt2000}). 
Since the ultraviolet absorption is produced by a resonance line while the X-ray 
 absorption by bound-free edges, the optical-to-X-ray spectral index, 
 $\alpha_{OX}$ is expected to be 
 practically independent of the C~IV EW up to a column density corresponding to 
 $\tau_{bf}\sim1$.
Above this value, a very rapid decline in $\alpha_{OX}$ and a slow increase 
 in C~IV EW is expected. 
We calculate the indices following Strateva et al. (\cite{Strateva2005}), 
 taking the fluxes at 2500${\AA}$ and 2 keV in the rest frame of the quasar,
  $\alpha_{OX} = \log \left( F_{2keV}/F_{2500\AA} \right) /  \log \left(\nu_{2keV} / \nu_{2500\AA}\right)$.
The correlation discovered between $\alpha_{OX}$ and C~IV EW instead exhibits  a 
 gradual increase in the C IV EW associated with a gradual decrease 
 in $\alpha_{OX}$ (compare Fig.~4 in Brandt et al. \cite{Brandt2000}).
This implies that either the UV and X-ray absorbers are distinct or 
 the X-ray weakness is produced mainly by a peculiar property of the nuclear 
 emission such as  extreme variability.  
In addition, Brandt et al. (\cite{Brandt2000}) noted several differences  
 between X-ray weak and ``normal'' quasars such as  low $[$O III$]$ luminosities 
 and equivalent widths, distinctive H$\beta$ profiles and a location 
 toward the weak  $[$O III$]$ end  of the Boroson \& Green eigenvector 1, 
 which may be indicative of a high  mass accretion rate relative to the Eddington limit.

Brandt et al. (\cite{Brandt2000}) speculated that with increasing 
 inclination angle we may observe quasars, X-ray weak quasars, 
 broad absorption line (BAL) quasars and type 2 quasars, which was 
also proposed by Elvis (\cite{Elvis2000}).
This idea has been supported by various observations. 
On the one hand, X-ray weak quasars were recognised as being BAL or mini-BAL quasars 
 (Gallagher et al. \cite{Gallagher2001}, Sulentic et al. \cite{Sulentic2006}). 
On the other hand, X-ray observations indicated high column densities for many X-ray 
 weak quasars (Gallagher et al. \cite{Gallagher2001}, \cite{Gallagher2004}, 
 Brinkmann et al. \cite{Brinkmann2004}, Piconcelli et al. \cite{Piconcelli2004}, 
 Schartel et al. \cite{Schartel2005}).

Various results challenge this picture and indicate greater complexity. 
In addition X-ray weakness is much less characteristic of BAL quasars class than previously understood 
 (Giustini et al. \cite{Giustini2008}, Blustin et al. \cite{Blustin2008}, Fan et al. \cite{Fan2009},
 Gibson et al. \cite{Gibson2009}). 
Some X-ray weak quasars exhibit significant variability being sometimes X-ray weak and sometimes not 
 (Ballo et al. \cite{Ballo2008}).
The variability is often correlated with a significant change in the spectral properties of the continuum emission.

Grupe et al. (\cite{Grupe2007}, \cite{Grupe2008b}) discussed whether varying absorbing column densities could explain the variable X-ray weakness and spectral properties found for \object{WPVS 007}. 
During the ROSAT All Sky Survey, this source was observed in an X-ray high state, but all subsequent observations  (between 1993 and 2007) detected fainter X-ray emission.
\object{WPVS 007} is particularly interesting because its FUSE spectrum obtained in 2003 exhibits a BAL outflow that is
 not present in the HST FOS spectrum acquired in 1996 (Leighly et al. \cite{Leighly2009}).
An X-ray weak continuum or even an  intrinsically X-ray weak spectrum in combination with the 
 X-ray absorption implied by the UV data provides the most probable interpretation of the observed 
 spectrum (Leighly et al. \cite{Leighly2009}).
Another X-ray weak quasar, \object{PHL 1811}, seems to be an intrinsically weak narrow-line 
 quasar (Leighly et al. \cite{Leighly2007a}).
Althought its X-ray spectral slopes are consistent with the values commonly observed in narrow-line Seyfert 1s,
 its X-ray spectra show no evidence of absorption and its UV spectra lack absorption lines  
 (Leighly et al. \cite{Leighly2007b}).
The narrow line quasar PHL 1092 exhibits a dramatic variability making it one of the most extreme 
 X-ray weak quasars (Miniutti et al. \cite{Miniutti2009}). 
The inferred steep power-law index and its persistent UV flux imply that an absorption event is unlikely, suggesting instead that a transient weakening or disruption of the X-ray emitting corona is responsible.

Schartel et al. (\cite{Schartel2007}) and Ballo et al. (\cite{Ballo2008}) discussed reflection\--dominated 
scenarios for low states of \object{PG~2112+059} and \object{PG~1535+547}, respectively.
Similar scenarios were  proposed by Gallo (\cite{Gallo2006}) for narrow-line Seyfert 1s, by Grupe et al. (\cite{Grupe2008a}) for the historical low state of 
 \object{Mrk 335}, and by  Vignali et al. (\cite{Vignali2008}) for the narrow-line quasar \object{PG 1543+489}.
The most impressive case of a reflection\--dominated AGN is the narrow-line Seyfert 1 \object{1H 0707-495}.
Its X-ray spectra show a sharp spectral drop, which Fabian et al. (\cite{Fabian2002}, \cite{Fabian2004}) 
 interpreted in terms of relativistically blurred ionised reflection from the accretion disk. 
A four\--orbit\--long XMM-Newton observation of  \object{1H 0707-495} in 2008 detected broad lines from ionised iron K and iron L.
The latter, observed for the first time in this type of observation, appears 20 times
 weaker than the iron K, in agreement with predictions based on atomic physics.
In addition, Fabian et al.    (\cite{Fabian2009}) measured for the first time the expected frequency-dependent time\--lags 
 between the power-law and the reflection component.
In spite of these developments, it is plausible that in some cases low states and X-ray weaknesses 
 are caused by suppressed continuum emission in the direction of our line of sight, as proposed by the light\--bending model  of Miniutti \& Fabian (\cite{Miniutti2004}).
In this model, the primary X-ray emitting source is located above the supermassive black hole and
 moves up and down.
If the source is within a few Schwarzschild radii of the black hole, then a substantial part of the 
 primary continuum emission may bend onto the accretion disk.
Therefore, observers see suppressed continuum emission and an enhanced reflection 
 component at the same time, which increases the probability of detecting emission from the accretion disk.

To study this possibility in more detail, we proposed a 200 ks deep X-ray observation of 
 \object{PG~2112+059} in a low or deep minimum state (Schartel et al. \cite{Schartel2007}). 
As the source is highly variable, we proposed to complete a short snapshot observation followed  by a targeted deep observation only if the source was detected at a flux level comparable to that measured in November 2005. 

In Sect.~\ref{Sum}, we characterise \object{PG~2112+059} and summarise its observation history in X-rays. 
The XMM-Newton observations and the subsequent data reduction are presented in Sect.~\ref{OaDR}.
 We shifted the observation and extraction tables to the Appendix for the convenience of our reader. 
The analysis of the timing behaviour of both the optical and X-ray data follows in Sect.~\ref{Var}. 
In Sect. \ref{SPE}, we provide the X-ray spectral analysis followed by the discussion in 
 Sect.~\ref{Dis} and our general conclusions in Sect.~\ref{Con}

\section{PG 2112$\rm+$059}
\label{Sum}

The quasar \object{PG~2112+059} was recognised to be such within the Palomar bright quasar survey 
 (Schmidt \& Green \cite{Schmidt1983}).
The source has a redshift of $z=0.456$ (V\'eron-Cetty \& V\'eron \cite{Veron2000}).
Based on the Hubble Space Telescope spectrum (Jannuzi et al. \cite{Jannuzi1998}), Gallagher et al. 
 (\cite{Gallagher2001}) could classify \object{PG~2112+059} as a BAL quasar.  
The supermassive black hole of PG 2112$+$059 has one of the highest masses of all Palomar quasars 
 with $\log M_{BH}/M_{solar}=9.0\pm0.1$ (Vestergaard \& Peterson \cite{Vestergaard2006}).

 \object{PG~2112+059} was observed by ROSAT in 1991, by ASCA in October 1999 
 (Gallagher et al. \cite{Gallagher2001}), by Chandra in September 2002 (Gallagher et al. \cite{Gallagher2004}) ,
 and twice by XMM-Newton in May 2003 and November 2005 (Schartel et al. \cite{Schartel2005}, \cite{Schartel2007}).
Based on the ROSAT data, Wang et al. (\cite{Wang1996}) first measured a low optical-to-X-ray 
 spectral index as is characteristic of X-ray weak quasars. 
The observations found dramatic variability in the X-ray energy range.
The highest flux was that detected by ASCA, which was approximately a factor of four higher than in the earlier 
 ROSAT observation as well as in the later Chandra observation.
In May 2003, the source was in an intermediate flux state and in 
 November 2005 a deep minimum state.

The ASCA spectrum of \object{PG~2112+059} detected clear evidence of absorption but the photon statistics
  were insufficient for any accurate characterisation.  
The Chandra spectrum is incompatible with neutral absorption completely covering the source.  
The spectrum can be described statistically assuming an ionised absorber as well as a partially covering 
 neutral absorber (Gallagher et al. \cite{Gallagher2004}),  where both absorption scenarios require an 
 increase in the absorbing column  density between 1999 and 2002.
Schartel et al. (\cite{Schartel2007}) interpreted the two XMM-Newton observations in the context of light 
 blending (Miniutti \& Fabian \cite{Miniutti2004}) but could not exclude complex absorption scenarios 
 involving two layers of absorbing  material with different ionisation states.

\begin{table*}[thb]
\caption[]{\label{VAROM2} Optical variability of PG 2112+059: OM fluxes}
\begin{tabular}{lccccccc}
\\
\hline
\\
\multicolumn{1}{c}{Filter} & \multicolumn{1}{c}{Effective} &\multicolumn{4}{c}{Epoch} &
\multicolumn{1}{c}{Average flux}   & \multicolumn{1}{c}{Ratio} \\
\multicolumn{1}{c}{}        & \multicolumn{1}{c}{Wavelength}
& \multicolumn{1}{c}{2003-05}
& \multicolumn{1}{c}{2005-11}
& \multicolumn{1}{c}{2007-05}
& \multicolumn{1}{c}{2007-11}
& \multicolumn{1}{c}{} & \multicolumn{1}{c}{} \\
\multicolumn{1}{c}{}
& \multicolumn{1}{c}{[nm]}
& \multicolumn{1}{c}{(1,2)}
& \multicolumn{1}{c}{(1)}
& \multicolumn{1}{c}{(1)}
& \multicolumn{1}{c}{(1)}
& \multicolumn{1}{c}{(1,3)}
& \multicolumn{1}{c}{(4)}
\\
\hline\hline
\\
V      & 500 - 590 & 22  $\pm$2   & 20.7$\pm$0.1 & 20.0$\pm$0.1 & 19.4$\pm$0.2 & 20.1$\pm$0.1 & 1.07$\pm$0.02 \\
B      & 380 - 490 & 31  $\pm$2   & 29.3$\pm$0.1 & 28.0$\pm$0.1 & 27.9$\pm$0.2 & 28.3$\pm$0.1 & 1.05$\pm$0.01\\
U      & 300 - 390 & 37  $\pm$3   & 38.9$\pm$0.1 & 36.4$\pm$0.1 & 35.4$\pm$0.2 & 36.9$\pm$0.1 & 1.10$\pm$0.01 \\
UVW1   & 240 - 360 &              & 46.7$\pm$0.2 & 43.4$\pm$0.1 & 42.4$\pm$0.2 & 43.7$\pm$0.1 & 1.10$\pm$0.01 \\
UVM2   & 200 - 270 &              & 60.7$\pm$0.8 & 55.0$\pm$0.4 & 54.3$\pm$0.7 & 55.7$\pm$0.3 & 1.12$\pm$0.03 \\
UVW2   & 180 - 230 &              & 63.8$\pm$1.3 & 56.6$\pm$0.8 & 54.9$\pm$1.2 & 57.5$\pm$0.6 & 1.16$\pm$0.05 \\
\\
\hline
\\
\multicolumn{6}{l}{(1) Flux in units of 10$^{-16}$erg/cm$^{2}$/s/{\AA}}\\
\multicolumn{6}{l}{(2) Estimated based on optical grism observation }\\
\multicolumn{6}{l}{(3) Average flux excluding 2003-05 data }\\
\multicolumn{6}{l}{(4) Maximum to minimum flux ratio excluding 2003-05 data  }\\
\end{tabular}
\end{table*}

\section{Observations and data reduction}
\label{OaDR}

In 2007, XMM-Newton (Jansen et al. \cite{Jansen2001}) observed 
 \object{PG~2112+059} four times. 
First, to determine the current state of the quasar, a short ($\rm \sim$25 ks) 
 snapshot-type observation (observation identifier (ObsId) 5005006) 
 was performed on the 3 May 2007. 
Since this observation detected the quasar in the targeted low state, 
 XMM-Newton pointed towards \object{PG~2112+059} for two adjoining 
 revolutions with the maximum possible exposure time of $\rm \sim$100 ks each from 
 the 19 May 2007 onwards (ObsId 5005007 and 5005008). 
About half a year later, during the autumn visibility window of the 
 target, a fourth observation (ObsId: 5005009) was performed
 to test whether the quasar remained in the low state. 
This observation started on the 5 November and accumulated 
 photons for $\rm ~$53ks. 

To take full advantage of the calibration progress and allow 
 an unbiased comparison of all XMM-Newton observations of \object{PG~2112+059},
 we re-extracted the data taken in May 2003 (ObsId:  150610201) and the 
 data taken in November 2005 (ObsId: 3003102). 
Details of these observations are given in Schartel et al. (\cite{Schartel2005}, \cite{Schartel2007}). 
In addition we considered the spectrum taken by Chandra in September 2002 (ObsId 3011). 
 Gallagher et al. (\cite{Gallagher2004}) provide details of this observation.

XMM-Newton carries three scientific instruments that observe simultaneously: 
 the Reflection Grating Spectrometer (RGS; Brinkman et al. \cite{Brinkman2001}), 
 the European Photon Imaging Camera (EPIC), and 
 the Optical Monitor (OM; Mason et al. \cite{Mason2001}). 
EPIC consists of three CCD cameras: 
  the pn-camera  (Str\"uder et al. \cite{Strueder2001}) and 
  two MOS-cameras (Turner et al. \cite{Turner2001}).  
All EPIC exposures of \object{PG~2112+059} were taken in full frame mode 
 with different optical blocking filters in the light path. 
OM observed in  an ``science user defined'' image mode with different filters or 
 grisms in the optical light path.  
The details of the exposures are provided in Appendix \ref{A1}.

We processed the data with the XMM-Newton Science Analysis System   
 (SAS) v. 7.1 (linux; compare Loiseau et al. \cite{Loiseau2007}) with the 
 calibration files generated between August and November 2007.
Neither the pipeline products provided by the Survey Science Centre 
 (Watson et al \cite{Watson2001}) nor our visual inspection of the processed 
 RGS events in the dispersion-cross-dispersion-plane allowed a detection of 
 the source with RGS.

We processed the EPIC and OM data exactly as described in Schartel et al. 
 (\cite{Schartel2007}), applying the screening procedure for time ranges 
 with low radiation background level according to 
 Piconcelli et al. (\cite{Piconcelli2005}). 
The details of the extraction and its results are described in 
 Appendix \ref{A2}.

The {\it Chandra} data reduction was performed using the Ciao 3.4 
 version applying the calibration according to CALDB 3.4.0 (May 2007).  
Source and background spectra, and response matrices were 
 produced using the {\it psextract ciao} task.   
The source spectrum  was  extracted from  a circular  region 
 centred  on the maximum  emission with a  4.92$\rm \arcsec$ (10  pixels)  
 radius. 
The background spectrum was extracted using five different regions 
 free of any contaminating source. 
They are all located in the same CCD with a distance  
 of 25-40$\rm \arcsec$ to the source. 
The {\it Chandra} spectrum was binned to ensure at least 20 
 counts per energy bin.

\section{Timing analysis}
\label{Var}

\subsection{OM}

To test for optical variability on short time scales, we face the difficulty that within 
 each observation only a few exposures for each of the optical filters were performed. 
Therefore, for each filter we divided the maximal flux by the minimal one and propagated 
 the errors accordingly. 
All results are in agreement with 1 within the errors.
Consequently, we exclude optical variability being present in the individual observations.

Table ~\ref{VAROM2} gives the mean fluxes as measured with the different OM filters for May 2003, 
 Nov 2005, and May and Nov 2007. 
{\rm It also provides per filter the ratio of the maximum
to minimum flux observed between 2005 and 2007 (i.e., excluding 2003 data),
as well as the average flux.}
The optical fluxes measured in 2007 are significantly lower than in 2005, the amplitude of the 
 variation being larger at shorter wavelengths.  
The fluxes also appear to decrease between May and November 2007, although the statistical evidence is much weaker. 
The ratios of maximum to minimum fluxes demonstrate that the variation correlates with wavelength.

\begin{figure}[t]
\centering
\includegraphics[width=6.5cm,angle=-90]{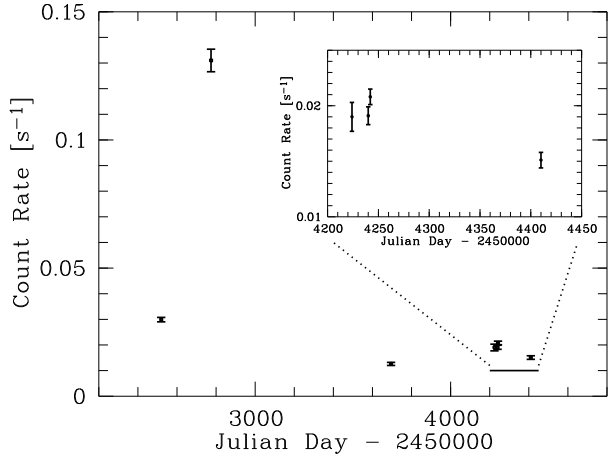}
\caption{ 
X-ray variability of PG 2112+059 between September 2002 and November 2007 is shown. 
The image shows the count rate measured by the pn camera as a function of the Julian date. 
The count rate measured with Chandra in 2002 was converted 
to a corresponding pn count rate (first data point). 
The inlay shows the count rates collected in 2007. }
 \label{time}
   \end{figure}

During the first XMM-Newton observation, performed in May 2003, only OM grism exposures were performed. 
To estimate the broad-band flux, we folded the flux observed in the optical grism through the effective 
 area of the filters. 
The results are provided in Table~\ref{VAROM2}. 
The quoted fluxes are uncertain and the errors are larger than the quoted statistical extraction errors.
The presence of spectra from other sources complicates the extraction of the quasar grism spectrum and 
 the result still depends on the selected source and background extraction regions. 
So, the maximal conclusion from the optical grism in 2003 is that the flux in that epoch did not differ to 
 within 10-20\% from the flux measured in later observations.

\subsection{X-ray}

To test for X-ray variability within the individual observations, we 
 created light curves for different time bins and energy ranges for 
 each of the four \object{PG~2112+059} observations performed in 2007. 
The light curves were created from the unscreened events collected 
 by the pn-camera as this camera has the largest effective area of all
  X-ray observing instruments   of XMM-Newton.
We extracted the source and background photons from the same sky area 
 as used for the spectral analysis (specified in the Appendix, 
 Table~\ref{TSCR}, for each observation).  
We produced light curves with time bins of 5000s, 10~000s, 15~000s and 20~000s 
 for the 0.35 to 0.5 keV, 0.5 to 2.0 keV, 2 to 10 keV, and 0.3 to 12 keV 
 energy bands. 
The softest band in combination with 5000 s or 10~000 s bin size shows too 
 few counts per bin to permit further analysis.
Inspecting the light curves, we found no obvious variability pattern and we 
 estimate the variability to be within a range of 20\%.

PG 2112+059 shows a pronounced variability over longer time scales. 
The pn count rates of the observations discussed in this paper are plotted 
 in Fig.~\ref{time}, where the count rate observed by Chandra was converted 
 based on a spectral fit. 
The count rates corresponding to the observation taken in 2007 are shown in 
 detail in the inlay. 
After the 2003 high state, PG 2112+059 has always been observed during low states
 showing a flux a factor from 7 to 10 lower than in a high state. 
The flux before the high state is about 30\% higher than the flux observed after 
 it. 
However, it should be noted that high states in\--between the measurements cannot be excluded.

\begin{figure}[t]
\centering
\includegraphics[width=9.5cm,angle=0]{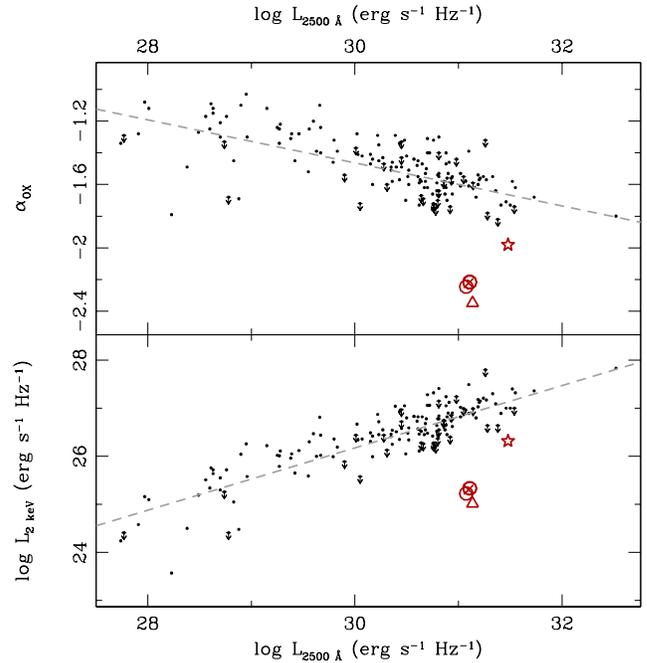}
\caption{ 
The dependence on the 2500$\AA$  monochromatic luminosity of $\alpha_{OX}$
 (upper panel) and L$_{2keV}$ (lower panel) is shown in comparison with the 
 values obtained for \object{PG~2112+059}.
Black-filled circles mark data from Strateva et al. (2005) for their 
 ``main'' SDSS sample (155 objects with 0.1 $ \lesssim $ z $ \lesssim $ 4.5); 
 arrows indicate upper limits to the X-ray detection. 
Dashed lines are the best-fit linear relations for their combined sample 
 (``main'' SDSS sample plus high-redshift sample plus Seyfert 1 sample).
The values derived for \object{PG~2112+059} are plotted in grey 
 (red in the colour edition).
The star labels the observation from May 2003, the triangle from November 2005, 
 the circles from May 2007, and the ``x'' from November 2007.
}
 \label{alphaox}
   \end{figure}

\subsection{$\alpha_{OX}$}

As mentioned, we calculated  $\alpha_{OX}$ following Strateva et al. (\cite{Strateva2005}) by taking 
 the fluxes at 2500$ \AA$ and 2 keV in the rest frame of the quasar.
%  $\alpha_{OX} = \log \left( F_{2keV}/F_{2500\AA} \right) /  \log \left(\nu_{2keV} / \nu_{2500\AA}\right)$. 
The X-ray flux was determined using fit 4 provided in Table~\ref{X_FITS_BLC}.
Both the X-ray and optical data were corrected for Galactic absorption. 
For the optical data, we applied the extinction curve provided by 
 Cardelli et al. (\cite{Cardelli1989}).
We did not correct the optical data for the contribution of the host galaxy, 
 thus the nuclear UV flux could be lower than the value estimated here.

The relation between rest-frame UV and soft X-ray AGN emission, and its dependence 
 on redshift and/or optical luminosity was investigated several times. 
Most studies detected no evidence of a redshift dependence, while the X-ray emission was found to be  
 correlated with the UV emission such that $\alpha_{OX}$ decreases as the UV emission 
 increases (e.g., Vignali et al. \cite{Vignali2003}, 
 Strateva et al. \cite{Strateva2005}, Steffen et al. \cite{Steffen2006}, 
 Just et al. \cite{Just2007}). 
Strateva et al. (2005) combined radio quiet sources from the SDSS, a 
 heterogeneous low-redshift Seyfert 1 sample, and a heterogeneous high-redshift sample. 
They found a correlation between the monochromatic luminosities at 2500 {\AA} and 2 keV. 
The broadband spectral index
is anticorrelated with the rest-frame monochromatic UV luminosity.
 \object{PG~2112+059} appears to be under-luminous in the X-ray band compared to
 optically selected sources in the same bin of UV luminosity.
In Fig.~\ref{alphaox}, we compare the UV and X-ray luminosities, and $\alpha_{OX}$ with the trend 
 found by Strateva et al. (\cite{Strateva2005}) for their sample of 155 optically 
 selected active galactic nuclei.

\begin{figure}[h]
\centering
\includegraphics[width=6.5cm,angle=-90]{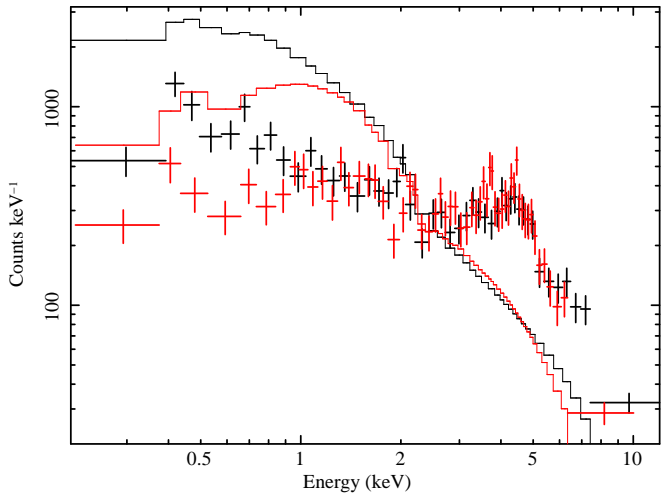}
\caption{  
The figure shows the observed (count) spectra of pn (black) and combined MOS 
 (grey \-- red in the colour edition) of PG 2112+059 taken in May 2007. 
In addition, we show the counts expected for a canonical power\--law model 
 ($\Gamma=1.89$) absorbed by Galactic absorption with arbitrary normalisation.
In comparison to the expectation for a canonical power law, the observed 
 spectra is very hard, which may be indicative of either a huge deficit below 5$~$keV 
 or a large excess emission in the 2 to 5 keV energy range.
  }
 \label{FigInt}
   \end{figure}

We also determined the absorption-corrected optical-to-X-ray spectral index. 
We based our calculation on the most accurately estimated parameters of fit 3 and  4 
 (Table\ref{X_FITS_BLC}).
The resulting values of $\alpha_{OX} < -1.7$ imply that \object{PG~2112+059} 
 was observed while intrinsically X-ray weak in each of the XMM-Newton observations.

\section{X-ray spectral analysis}
\label{SPE}

We performed the spectral analysis as described by Schartel et al. (2007).
Following the standard calibration recommendations (Guainazzi \cite{Guainazzi2009}), we 
 analysed the MOS spectra over the energy range from 0.2 keV to 10.0 keV and 
 the pn spectra over the 0.2 to 12.0 keV energy range.

The count rates of all three observations from May 2007 agree with those of a constant 
source, and a visual inspection showed no indication of
 differences in the spectral slope.
We therefore added the spectra obtained for each 
 instrument and calculated the corresponding effective areas and detector 
 response matrices giving weight according to the accumulated 
 exposure time.
In addition, we added the spectra obtained for MOS1 and MOS2 for all 
 observations and calculated the corresponding auxiliary files by applying 
 the same weighting with exposure time. 
As reported in Schartel et al. (2007) for the 2005 observation, 
 in all cases the accumulated counts at low energies were too 
 few to take advantage of the differences in effective area associated 
 with the different optical blocking filters.

\begin{table}[b]
\caption[]{\label{Mod} Spectral models }
\begin{tabular}{lll} \hline \\
Model & Description \& References & XSPEC   \\  
label &                          & Name     \\ \\ \hline \hline \\
 $i$& partial covering with ionised absorber at the              & zxipcf \\
    & redshift of the source (Reeves et al. \cite{Reeves2008})   &         \\ 
 $d$& reflection by distant neutral material                  & pexmon   \\ 
    & including fluorescence emission lines                   &         \\
    & (Nandra et al. \cite{Nandra2007})                      & \\ 
 $p$& power\--law                                             & pow       \\

 $k$& relativistic blurring from an accretion disk            & kyconv   \\
    & around a rotating Kerr  black hole                      &           \\
    & (Do\v{v}ciak et al.\cite{Dovciak2004})                 &           \\  
 $n$& neutral absorber at the redshift of the source          & zphabs   \\ 
 $r$& reflection by a constant density illuminated            & reflion  \\ 
    & atmosphere (Ross \& Fabian \cite{Ross2005})            &  \\ \\
\hline \\ 
\end{tabular}
\end{table}

Reflecting the high number of collected counts, we binned the May 2007 
 combined pn spectrum to achieve a signal-to-noise ratio of at least 6. 
All other pn and combined MOS spectra were binned to reach a signal-to-noise 
 ratio of at least 5. 
In all cases, we limit the binning to three bins per spectral resolution element.
We used XSPEC 12.5.0ac (Linux version, Arnaud \cite{Arnaud2009})  in the 
 spectral analysis that estimates spectral parameters with the modified 
 minimum $\chi^2$ method (Kendall and Stuart \cite{Kendall1973}). 
Errors in estimated parameters are provided for the 90\% confidence region 
 for a single interesting parameter   ($\Delta \chi^2 = 2.7$, Avni 
 \cite{Avni1976}).  
All modelling was performed based the assumption of Galactic foreground 
 absorption with an equivalent column density of 
  $\rm N_H = 6.22 \times 10^{20} \rm cm^{-2}$  
 (Kalberla et al. \cite{Kalberla2005}). 
For the plasma emission and photoelectric absorption, we assumed solar 
 abundances as determined by Anders and Ebihara (\cite{Anders1982}) and a
 photoelectric cross\--section according to Verner et al. (\cite{Verner1996}).
We assumed a cosmology with $\rm \Omega_M = 0.27$, $ \rm \Omega_{\Lambda} = 0.73$ 
 and $\rm H_0 = 70 \; km s^{-1} Mpc^{-1}$ (Spergel \cite{Spergel2007}).

The signal-to-noise ratios obtained for the November 2007 spectra are too 
 low to allow us to set constraints of the physical emission mechanisms.
We therefore focus in the following on the added spectra of the May 2007 
 observation and use the November spectra only to check for consistency 
 and in the joint modelling of all the obtained spectra.

\begin{table*}
\caption[]{\label{FITS_IOC} Spectral fits of the  May 2007 EPIC spectra of \object{PG~2112+059}: Continuum properties }
\begin{tabular}{llccccccrr}
\hline \\ 
Fit & Model$^{(a)}$     
    & \multicolumn{3}{c}{Absorber}      
    & \multicolumn{2}{c}{Continuum} 
    & \multicolumn{1}{c}{Distant}  
    & \multicolumn{2}{c}{Statistic}  \\ 
    &                
    & \multicolumn{3}{c}{neutral}       
    & \multicolumn{2}{c}{power-law}
    & \multicolumn{1}{c}{Reflection} 
    & \multicolumn{2}{c}{ }  \\ 
    &                   
    & \multicolumn{1}{c}{$N_H^{(b)}$}    & \multicolumn{1}{c}{$cf^{(c)}$}  & \multicolumn{1}{l}{$\xi^{(d)}$}  
    & \multicolumn{1}{c}{$\Gamma^{(e)}$} & \multicolumn{1}{c}{$Norm^{(g)}$} 
    & \multicolumn{1}{c}{$Norm^{(g)}$} 
    & \multicolumn{1}{c}{$\chi^2$}     & \multicolumn{1}{c}{d$^{(h)}$}   \\ 
    &                   
    & \multicolumn{1}{l}{{\small [$\rm 10^{22} cm^{-2}$]}}    & \multicolumn{1}{l}{{\small }}  & \multicolumn{1}{l}{{\small [$ \log$]} }  
    &  & \multicolumn{1}{r}{\small{$[\rm  keV^{-1}cm^{-2}s^{-1}]$}}
    & \multicolumn{1}{l}{\small{$[\rm  keV^{-1}cm^{-2}s^{-1}]$}} 
    & \multicolumn{1}{c}{} & \multicolumn{1}{c}{}   \\ \\ 
\hline \hline \\  
1   &  $i * (p + d) $             
    &  $18.3_{-1.9}^{+2.7}$   & $0.94_{-0.02}^{+0.01}$           & $1.10_{-0.52}^{+0.45}$  
    &  $2.08_{-0.13}^{+0.16}$ & $8.74_{-2.51}^{+2.29}\; 10^{-5}$
    & $9.24_{-5.14}^{+8.03}\; 10^{-5}$
    &  93.2         & 99            \\ \\
%=============================================================
2   &  $i * (p + k(r) + d) $             
    &  $0.12_{-0.02}^{+0.39}$ & $1.00_{-0.14}^{+0.00}$           & $-1.36_{-0.31}^{+3.04}$  
    &  $1.53_{-0.10}^{+0.16}$ & $3.58_{-0.89}^{+0.74}\; 10^{-6}$
    & $1.45_{-0.95}^{+1.17}\; 10^{-5}$
    &  96.7         & 93            \\ \\

\hline \\ 
\end{tabular}
                                                       
Explanation of labels: 
 (a) the spectral models described in Table~\ref{Mod},
 (b) equivalent column density, 
 (c) covering fraction,
 (d) ionisation level of absorbing material $\xi=L/nr^{2}$,  as defined in {\em XSPEC},
 (e) photon index of power-law continuum, 
% (f) fixed,
 (g) normalisation of photon flux at 1 keV, 
 (h) degrees of freedom.
The best\--fit parameters for the ionised reflector, the disk, and the black hole 
 are provided in Table \ref{FITS_IOL}. 
All errors are provided at the 90\% confidence level.
\end{table*}

\begin{table*}
\caption[]{\label{FITS_IOL}   Spectral fits of the  May 2007 EPIC spectra 
of \object{PG~2112+059}: Ionised reflector, disk, and black hole properties}
\begin{tabular}{llcccccc}
\hline \\ 
Fit & M$^{(a)}$& \multicolumn{3}{c}{Ionised Reflector} &   \multicolumn{2}{c}{ Disk} &   \multicolumn{1}{c}{Black Hole}   \\
    &  & Norm         &  $Fe^{(b)}$ & $\Xi^{(c)}$  &   Index$^{(d)}$ & Inclination    &  $a/M^{(e)}$ \\ 
    &         & [$ \rm cm^{-2}\;s^{-1}$]  &   [$\rm erg \, cm \,s^{-1}$] &        & [$\rm degree$] &        \\ \\
\hline \hline \\ \\
2 &k(r)&$1.69_{-1.25}^{+11.7}\;10^{-6}$&  $1.74_{-0.30}^{+0.32}$ &   $30.0_{-0.0}^{+17.8}$ &$3.29_{-0.58}^{+2.00}$&$52.1_{-3.0}^{+3.1}$ &$0.93_{-0.10}^{+0.07}$ \\
\\
\hline \\
\end{tabular}

The corresponding continuum properties are provided in Table \ref{FITS_IOC}.
Explanation of labels: 
 (a) the spectral models described in Table~\ref{Mod},
 (b) iron abundance relative to solar iron abundance,
 (c) ionisation parameter as defined in {\em  reflion},
 (d) power\--law dependence of emissivity, and
 (e) black hole angular momentum.
All errors are provided at the 90\% confidence level.
\end{table*}

\subsection{Spectra taken in May 2007}
\label{May2007}

For the spectral analysis, we fitted the pn spectrum and the MOS spectrum, 
obtained by adding the MOS1 and MOS2 spectra,  together constraining all the free parameters to have the same value for the two 
 data sets. 
Reflecting differences in the absolute calibration of the two instruments 
 (Guainazzi \cite{Guainazzi2009}), the model for the MOS spectrum was 
 multiplied by 1.07. 

As for the spectra measured during observations in November 2005 
 (Schartel et al. \cite{Schartel2007}), the spectra of PG 2112$+$059 taken 
 in May 2007  is rather hard without showing a pronounced narrow iron 
 K$\alpha$ fluorescent emission line as expected for heavily absorbed sources. 
For illustration, we show in Fig.~\ref{FigInt} the pn and MOS count
 spectra, and for comparison the counts expected for a canonical power\--law 
 model ($\Gamma=1.89$, Piconcelli et al. \cite{Piconcelli2005}) absorbed 
 by Galactic absorption with arbitrary normalisation.
A canonical power\--law model may provide a first-order good approximation 
 for energies $>$5 keV, implying an enormous  deficit at lower energies.
In addition a canonical power\--law may provide a first-order good approximation 
 of the soft energy range (0.2 to 2 keV)  that fails in this case to account for 
 the hard band: in particular, the data show a large excess compared with this model in the 2 to 5 keV energy range.
Starting from these considerations and taking into account the results obtained from 
 previous observations, we attempt to explain the spectra within an 
 absorption scenario and compare the results with a model based on reflection within an ionised disk.

\subsubsection{Simple absorption models}
\label{SAM}

We cannot describe the spectra with a power law source observed behind a layer of neutral 
 ($\rm \chi^2=378.4$, $\rm d.o.f. = 102$) or ionised 
 ($\rm \chi^2=158.5$, $\rm d.o.f.=101$) material located at the redshift of the source. 
Within the analysis, two descriptions of the ionised absorber were used as 
 provided in XSPEC by the models {\it absori}\footnote{Ionised absorber model 
 (Zdziarski et al. \cite{Zdziarski1995}); it is just a first approximation, the temperature being considered 
 as an input parameter, and not calculated self-consistently by proper thermal balance.}
 and {\it zxipcf}\footnote{Model for partial covering of a partially ionised absorbing 
 material, based on a pre-calculated grid of XSTAR photoionised absorption model (Reeves et al. \cite{Reeves2008}).}.
To keep the discussion simple, here and in the following we refer to results 
 obtained with {\it zxipcf} (see Table~\ref{Mod}) only.
A partially covering neutral absorber allows a marginal statistically acceptable description of 
 the spectra ($\rm \chi^2 = 129.2$, $\rm d.o.f. = 101 $),  but shows systematic residua 
 in the 3 to 6 keV energy range.

\subsubsection{Complex absorption models and distant reflection}

\begin{figure}[h]
\centering
\includegraphics[width=6.5cm,angle=-90]{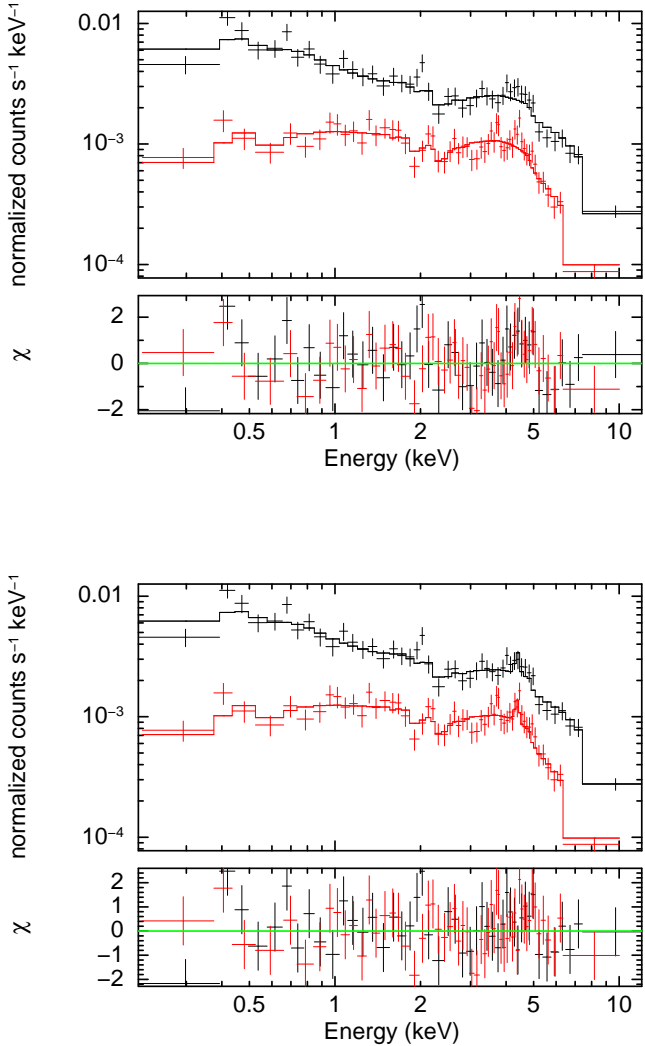}
\caption{
The figure shows the two EPIC spectra of PG 2112+059 taken in May 2007.
 The black data points were taken with the pn camera, the grey (red in colour edition) with the MOSs. 
The spectra are modelled with a power--law continuum modified by a partially covering ionised absorber.
Bottom panel shows the spectra in comparison to fit 1 listed in Table~\ref{FITS_IOC}.
It also includes a cold reflection component (continuum plus Fe K $\alpha$ line) parameterised by the pexmon model (Nandra \cite{Nandra2007}).
}
 \label{Fig1}
   \end{figure}

We model the spectra assuming various combinations of different absorbers. 
We considered neutral and ionised material, both totally or partially covering the source.
From the statistical viewpoint, all models provide an acceptable description of the spectra, 
 but exhibit a feature at  4.4 keV in the observers frame. 
The corresponding energy in the rest\--frame of the quasar, 6.4 keV, equals exactly 
 the energy of the K$_{\alpha}$ fluorescence line of neutral iron.
To test for the presence of an iron line, we focus on one spectral model: a power\--law 
 continuum observed behind ionised material partially covering the primary X-ray emitting region. 
In Fig.~\ref{Fig1}, we compare the data to the model (upper panel). 
We added to the spectral model a narrow Gaussian line with the energy fixed at 6.4 keV in 
 the rest frame of the quasar of fixed width 10 eV.
The $\chi^2$ decreased by  $\Delta \chi^2 = 10.8$  with one additional free parameter. 
Being aware of the problem of  applying statistical test in this specific context 
 (Protassov et al. \cite{Protassov2002}), an F-test  formally infers a probability of  
  1.1\% that the improvement reflects random chance.  
We conclude that neutral material reflecting X-rays in the direction of our line of sight is present.  

To consistently describe the neutral iron fluorescence emission line and the
associated reflection (contributing to the broad band, and specifically to the
2-5 keV bump), we added a reflection component to the the fit.
We modelled the reflection following Nandra et al. (\cite{Nandra2007}). 
In this approach, the narrow line and the Compton reflection continuum are handled 
 in a self-consistent way. 
The model is based on the neutral reflection model of Magdziarz \& Zdziarski (\cite{ Magdziarz1995}),
 which assumes a slab geometry, and on the emission\--line equivalent\--width simulations 
 of  George \& Fabian (\cite{George1991}). 
In addition, the model considers the Compton shoulder according to the description of 
 Matt (\cite{Matt2002}), and the Fe K$\beta$ and  Ni K$\alpha$ 
 lines with a flux of 11.3 per cent and  5 per cent of the iron K$\alpha$, respectively. 
The model can be used within XSPEC under the name {\it pexmon}. 
Since we are unable to differentiate the reflection component from the continuum emission, we must minimise the number of free parameters, leaving only the reflection normalisation 
 free to vary. 
Therefore, for all spectral modelling presented in this paper we fixed the {\it pexmon} photon index 
 to the photon index of the power\--law continuum.
In addition we fixed the cut-off energy to 150 keV, the iron 
 abundance 
\begin{center}
\includegraphics[width=6.5cm,angle=-90]{12389f04.ps}
\end{center}
to solar abundance, and the inclination angle to 45 degrees. 
The model was run such that only the reflected component was given, i.e., we set the relative 
 reflectivity to be -1.

We repeated the modelling of the spectra with the complex absorption scenarios assuming 
 additional reflection by neutral material located far away from the X-ray emitting source. 
All models provide acceptable descriptions of the data and exhibit a  $\rm \chi^2$-value lower 
 than found without reflection, where $\rm \Delta \chi^2$ ranges from 
 $\rm \Delta \chi^2 = 18.5$ to $\rm \Delta \chi^2 = 7.2$. 
About 10\% to 20\% of the ``bump'' emission can 
be attributed to the reflection component.
The results for a partially covering warm absorber model are listed in Table~\ref{FITS_IOC} 
 (fit 1), and in the lower panel of Fig.~\ref{Fig1} we compare the data  to the model.

\begin{figure}[h]
\centering
\includegraphics[width=6.5cm,angle=-90]{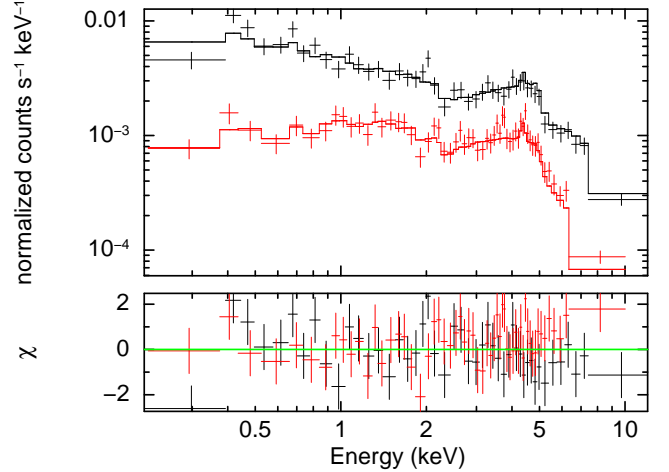}
\caption{
The figure shows the pn (black) and MOS (grey,- red in colour edition) spectra of 
 PG 2112+059 from May 2007 in comparison with a model (upper panel) and the 
 corresponding  residua (lower panel).
The model assumes a power\--law continuum, reflection on ionised material  
 (Ross \& Fabian \cite{Ross2005}) located in an accretion disk and reflection 
 on neutron material far away from the primary X-ray emitting source 
 (Nandra et al. \cite{Nandra2007}). 
The blurring was done following D\v{o}vciak, Karas \& Yaqoob (\cite{Dovciak2004}). 
All three components are assumed to be located  behind a ionised absorber. 
The best estimated parameters are provided in Table~\ref{FITS_IOC} and Table~\ref{FITS_IOL} under fit 2.
  }
 \label{Fig3}
   \end{figure}

\subsubsection{Reflection from an ionised accretion disk}
\label{BATL}

Following our previous results (Schartel \cite{Schartel2007}), we tested the possibility 
 of describing the spectra assuming X-ray ionised reflection by the accretion disk. 
Our model consists  of three additive components: 
 (1) a power--law continuum, which describes the primary X-ray emitting source; 
 (2) reflection by ionised material (Ross \& Fabian \cite{Ross2005}; 
     Crummy et al. \cite{Crummy2005}) blurred according to  
     D\v{o}vciak, Karas \& Yaqoob   (\cite{Dovciak2004}), 
     which describes the disk emission; and 
 (3) reflection by neutral material following Nandra et al. (\cite{Nandra2007}), 
     which allows us to account for possible reflection by material far away from the 
     supermassive black hole.
We assume that all three components are located behind a ionised, partially covering absorber.

To reduce the number of free parameters of the ionised reflection model, we tied the 
 iron abundance of the ionised reflector to the iron abundance of the neutral reflection
 and  
 the power\--law index of the reflection model to the index of the primary continuum. 
Two parameters of the reflection model were allowed to vary: the ionisation level of the 
 reflecting material and the normalisation.
The convolution model of D\v{o}vciak, Karas \& Yaqoob (\cite{Dovciak2004}) allows us to
  treat the black hole spin as a free parameter.
Haardt's limb brightening is taken into account and the calculations are performed for 
 1000 grid points in the local energy frame, without renormalising the reflection component. 

As we were unable to constrain the inner and outer radii of the emission region of the 
 accretion disk, we fixed the inner radius to the lowest marginally stable orbit and 
 the outer radius to $\rm r_{out} = \; 400 \;G M c^{-2} $. 
A trial without a neutral reflection component leaves residuals at the position of the 
 redshifted iron line, as we found for the absorption models in the previous section, and 
 therefore supports our model selection

The model allows a satisfying description of the data (e.g., fit2 in Table \ref{FITS_IOL}).
In Table~\ref{FITS_IOC}, we list the best\--fit values of the disk and black hole parameters derived by applying this model.
Figure~\ref{Fig3} compares the data  to the best\--fit model. 
The data are consistent with the assumption of a maximally rotating black hole.

\begin{table*}[th]
\caption[]{\label{X_FITS_BLC} Joint fits to the EPIC spectra of \object{PG~2112+059} taken at different epochs: continuum properties }
\begin{tabular}{r@{$\;\;$}rr@{ }r@{ }rr@{ }r@{ }rr@{ }rrr@{ }r}
\hline \\ 
Fit & Model$^{(a)}$     
    & \multicolumn{6}{c}{Absorber}      
    & \multicolumn{2}{c}{Continuum} 
    & \multicolumn{1}{c}{Distant}  
    & \multicolumn{2}{c}{Statistic}  \\ 
    &                
    & \multicolumn{3}{c}{warm}    
    & \multicolumn{3}{c}{warm}   
    & \multicolumn{2}{c}{power-law}
    & \multicolumn{1}{c}{Reflection} 
    & \multicolumn{2}{c}{ }  \\  
    &                    
    & \multicolumn{1}{c}{$N_H^{(b)}$}     & \multicolumn{1}{c}{$cf^{(c)}$}   & \multicolumn{1}{c}{$\xi^{(d)}$}
    & \multicolumn{1}{c}{$N_H^{(b)}$}     & \multicolumn{1}{c}{$cf^{(c)}$}   & \multicolumn{1}{c}{$\xi^{(d)}$}
    & \multicolumn{1}{c}{$\Gamma^{(e)}$}  & \multicolumn{1}{c}{$Norm^{(g)}$}        
    & \multicolumn{1}{c}{$Norm^{(g)}$} 
    & \multicolumn{1}{c}{$\chi^2$}        & \multicolumn{1}{c}{d$^{(h)}$}   \\                    
    &
    & \multicolumn{3}{c}{{\tiny [$\rm 10^{22} cm^{-2}$] $\;\;\;\;\;$ [$\log$]}}
    & \multicolumn{3}{c}{{\tiny [$\rm 10^{23} cm^{-2}$] $\;\;\;\;\;$ [$\log$]}}
    & \multicolumn{2}{r}{{\tiny{[$  keV^{-1}cm^{-2}s^{-1}$]}}}
    & \multicolumn{1}{r}{{\tiny{[$  keV^{-1}cm^{-2}s^{-1}$]}}}
    & \multicolumn{1}{c}{} & \multicolumn{1}{c}{}   \\ \\  \hline \hline \\ 
3   &  {\tiny $i * i * (p + d)$ }              
    & $4.98_{-1.05}^{+0.95}$ & $0.87_{-0.04}^{+0.03}$ &  
    & $3.47_{-1.09}^{+0.86}$ & $0.71_{-0.09}^{+0.08}$ & 
    & $2.22_{-0.15}^{+0.14}$ & 
    & $2.43_{-0.86}^{+1.22} \times 10^{-4}$    
    &  &  \\  
    &             
    & & & $+0.27_{-0.55}^{+0.40}$  
    & & & $+3.07_{-0.24}^{+0.35}$  
    & &   $9.40_{-1.80}^{+2.46} \times 10^{-5}$ 
    & & &  \\     
    &             
    & & & $+1.50_{-0.16}^{+0.30}$  
    & & & $+3.74_{-1.21}^{+2.26}$  
    & &   $1.98_{-0.44}^{+0.53} \times 10^{-4}$ 
    & & &  \\    
    &             
    & & & $-1.93_{-1.07}^{+0.99}$  
    & & & $+3.21_{-0.28}^{+0.55}$  
    & &   $3.89_{-0.92}^{+1.21} \times 10^{-5}$ 
    & & &  \\    
    &             
    & & & $-0.68_{-0.35}^{+0.77}$  
    & & & $+1.91_{-0.32}^{+0.09}$  
    & &   $1.24_{-0.32}^{+0.49} \times 10^{-4}$ 
    & & &  \\    
    &             
    & & & $-0.55_{-0.54}^{+0.97}$  
    & & & $+1.83_{-0.63}^{+0.22}$  
    & &   $8.70_{-2.27}^{+4.33} \times 10^{-5}$ 
    & &  256.8 & 243 \\    \\
4   &  {\tiny $i * (p + k(r) + d)$ }   
    & $0.25_{-0.08}^{+0.09}$ & $1.00^{f}$ &   
    &  &  &  
    & $1.65_{-0.05}^{+0.07}$ &  
    & $1.91_{-0.91}^{+1.03} \times 10^{-5}$    
    &  &  \\    
    &             
    & & & $-0.60_{-0.50}^{+0.47}$  
    & & &   
    & &   $ 6.37_{-6.37}^{+3.30} \times 10^{-6}$ 
    & & &  \\  
    &             
    & & & $+1.86_{-0.80}^{+0.95}$  
    & & &   
    & &   $ 5.97_{-0.67}^{+0.62} \times 10^{-5}$ 
    & & &  \\    
    &             
    & & & $-0.79_{-0.31}^{+0.38}$  
    & & &   
    & &   $ 0.00_{-0.00}^{+1.43}\times 10^{-6}$ 
    & & &  \\    
    &             
    & & & $-0.98_{-0.26}^{+0.36}$  
    & & &   
    & &   $ 3.27_{-1.90}^{+0.81} \times 10^{-6}$ 
    & & &  \\    
    &             
    & & & $-1.09_{-0.29}^{+0.37}$  
    & & &   
    & &   $ 2.13_{-2.13}^{+1.07} \times 10^{-6}$ 
    & &  265.3 & 234 \\    \\
\hline \\
\end{tabular}

The first line of each modelling provides the best\--fit model parameters of the joint fit.
The second to the sixth line provide only the parameters that were free to vary 
 between the spectra taken different epochs: September 2002 (Chandra observation of 
 low state, second line), May 2003 (high state, third line), November 2005 (deep 
 minimum state, fourth line), May 2007 (low state, fifth line), and November 2007 
 (low state, sixth line).
Explanation of labels: 
 (a) the spectral models are described in Table~\ref{Mod},
 (b) equivalent column density, 
 (c) covering fraction,
 (d) ionisation level of absorbing material $\xi=L/nr^{2}$,  as defined in {\em XSPEC},
 (e) photon index of power-law continuum, 
 (f) fixed,
 (g) normalisation of photon flux at 1 keV, 
 (h) degrees of freedom.
The best fitted parameters for the ionised reflector, the disk and the
 black hole are provided in Table \ref{X_FITS_BL}. 
All errors are provided for the 90\% confidence level.

\end{table*}
\begin{table*}
\caption[]{\label{X_FITS_BL}  Joint fits to the EPIC spectra of \object{PG~2112+059} taken at different epochs:
ionised reflector, disk, and black hole properties}
\begin{tabular}{llcccccc}
\hline \\ 
Fit & M$^{(a)}$& \multicolumn{3}{c}{Ionised Reflector} &   \multicolumn{2}{c}{ Disk} &   \multicolumn{1}{c}{Black Hole}   \\
    &  & Norm         &  $Fe^{(b)}$ & $\Xi^{(c)}$  &   Index$^{(d)}$ & Inclination    &  $a/M^{(e)}$ \\ 
    &         & [$ \rm cm^{-2}\;s^{-1}$]  &   [$\rm erg \, cm \,s^{-1}$] &        & [$\rm degree$] &        \\ \\
\hline \hline \\ \\
4 &k(r)& & $1.80_{-0.24}^{+0.26}$ & & & $52.5_{-2.3}^{+2.2}$ & $0.99_{-0.02}^{+0.0082}$ \\
  & & $8.35_{-5.08}^{+10.9}\;10^{-6}$ & & $31.2_{-1.2}^{+8.2}$  & $4.45_{-0.63}^{+0.92}$ & & \\
  & & $8.77_{-4.10}^{+2.70}\;10^{-7}$ & & $30.0_{-0.0}^{+23.1}$ & $3.00_{     }^{(f)    }$ & & \\
  & & $2.96_{-1.46}^{+2.91}\;10^{-7}$ & & $39.3_{-7.3}^{+ 9.7}$ & $2.87_{-0.21}^{+0.54}$ & & \\
  & & $6.55_{-2.06}^{+4.16}\;10^{-7}$ & & $30.0_{-0.0}^{+3.5}$  & $2.89_{-0.20}^{+20}$ & & \\
  & & $4.13_{-2.45}^{+4.63}\;10^{-7}$ & & $30.0_{-0.0}^{+7.2}$  & $2.79_{-0.39}^{+0.51}$ & & \\ \\

\hline \\
\end{tabular}

The corresponding continuum properties are provided in Table \ref{X_FITS_BLC}.
Explanation of labels: 
 (a) the spectral models are described in Table~\ref{Mod},
 (b) iron abundance relative to solar iron abundance
 (c) ionisation parameter as defined in {\em  reflion},
 (d) power\--law dependence of emissivity, 
 (e) black hole angular momentum,
 (f) fixed.
All errors are provided for the 90\% confidence level.
\end{table*}

\subsection{Spectra taken in November 2005}

Schartel et al.  (\cite{Schartel2007}) fitted only models of the disk assuming blurring 
 according to Laor  (\cite{Laor1991}). 
Therefore, to compare the two emission states, we modelled the EPIC spectra of PG 2112+059 
 taken in November 2005 assuming an ionised disk where the model is blurred according to
 D\v{o}vciak, Karas \& Yaqoob  (\cite{Dovciak2004}). 
Adding neutral reflection according to Nandra et al. (\cite{Nandra2007}) does not improve the description 
 of the data 
probably because of the poor statistics of the data. 
The statistics are also  not good enough to constrain the inclination angle,
 the data being consistent with the assumption of a maximally rotating black hole.

\subsection{Spectra taken in November 2007}
The spectra taken in November 2007 yield a poor quality statistics and cannot be 
 used to constrain the physical model. 
However,  they are consistent with the ionised disk 
 interpretation assuming blurring according to both Laor (\cite{Laor1991}) 
 or  D\v{o}vciak, Karas \& Yaqoob  (\cite{Dovciak2004}). 
The data are consistent with the assumption of a maximally rotating black hole.

\begin{figure*}[th]
 \centering
\includegraphics[width=13cm,angle=-90]{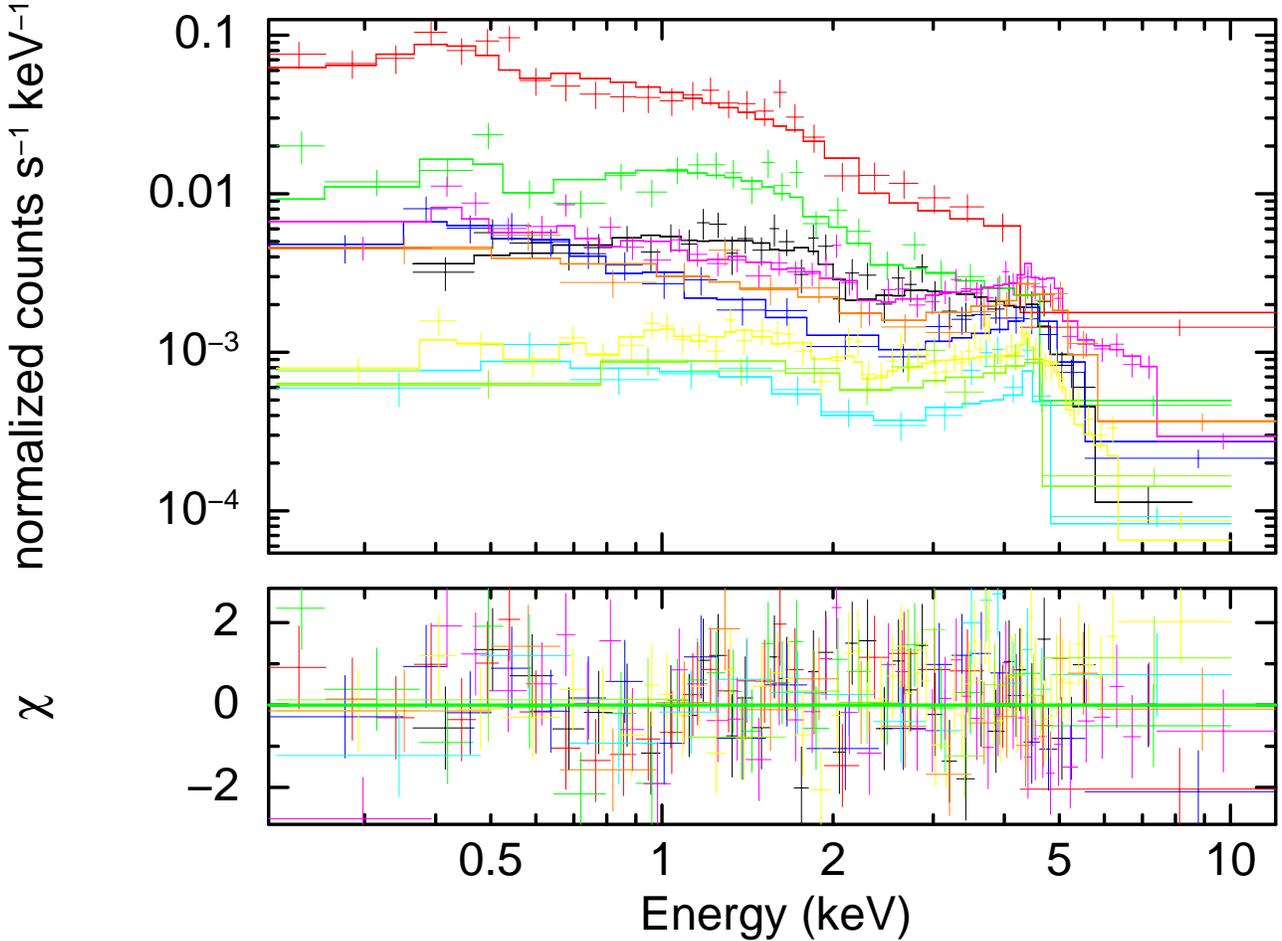}
\caption{The spectra of \object{PG~2112+059} at five different epochs 
(September 2002, intermediate state; May 2003, high state; November 2005, deep minimum state;  
 May 2007, low state; and November 2007, low state) in comparison to a model consisting of 
 a continuum power\--law and a relativistically blurred X-ray ionised reflection component 
  that originates in an accretion disk close to a supermassive  black hole. 
Both components are observed through a warm absorber and reflected by neutral distant material
 (Table~\ref{X_FITS_BLC} and \ref{X_FITS_BL}, fit 4). }
 \label{Fig4}
   \end{figure*}

\subsection{Joint fits to the spectra taken at different epochs}
From a mathematical viewpoint, the complex absorption models as well as the 
 models that assume reflection by an ionised disk allow a satisfying description of the data when the states are analysed separately. 
We therefore studied whether the different scenarios allow a consistent description 
 of the spectra observed at five different epochs. 
Since the source was observed in different states, our hope is that additional physical 
 arguments can be found.

For each epoch, we fitted the pn spectrum and the combined MOS spectrum 
 (obtained by adding the MOS1 and MOS2 spectra) in one fit, forcing all free and frozen 
 parameters to have the same value for both data sets. 
Reflecting differences in the absolute calibration (Guainazzi \cite{Guainazzi2009}), the model for the 
 MOS spectrum was multiplied with 1.07. 
All the models below include reflection from distant material, which was found to clearly be present.

\subsubsection{Absorption models}

A model consisting of both a power law and reflection by neutral material, where both components 
 are partially covered  by ionised material does not allow a satisfying description of the 
 data ($\chi^2=599.5$ for $d.o.f.=246$), although the covering fraction and the ionisation 
 parameter were allowed to vary. 
Adding a second layer of partially covering ionised material, we achieved a satisfying 
 description of the data. 
The results are provided in Table~\ref{X_FITS_BLC} as fit3.
We note that the description of the data by this model does not require 
 any change in the covering  fraction of the absorbing material.

In addition, we modelled the data allowing the covering fraction of both ionised 
 absorbers to vary for the different source states. 
We obtain a $\chi^2 = 234.5$ for $d.o.f. = 235$. 
An F-test shows that the model probably provides an improved description of the data, 
 but below the 3-$\sigma$ threshold.
The two absorption systems are characterised by $N_H \sim 5 \times 10^{22} cm^{-2}$ 
 and $N_H \sim 2.3 \times 10^{23} cm^{-2}$. 
The ionisation parameters of the two systems do not show any separation.  
Especially, it is impossible to differentiate between a high and low ionisation system. 
In addition, the ionisation parameters do not indicate any correlation with the 
 normalisation of the power\--law continuum. 
Therefore, we do not discuss this model any further in the following.

\subsubsection{Reflection from an ionised accretion disk}

As in the previous section, we now describe how we jointly modelled the five spectra taken at  
 different epochs assuming ionised reflection by the accretion disk that is blurred 
  according  to D\v{o}vciak, Karas \& Yaqoob (\cite{Dovciak2004}), (see 
 Sect.~\ref{BATL}), seen behind a layer of ionised material. 
We verified that the modelling improves significantly by adding a reflection component 
 from neutral material that is far away from the X-ray emitting source.  
The results of the analysis are provided in Tables \ref{X_FITS_BLC} and \ref{X_FITS_BL} as fit 4.

For the different epochs, the ionisation level of the warm absorber, the power\--law continuum 
 normalisation, the radial dependence of the emission of the disk, the ionisation level of 
 the ionised reflector, and the normalisation of the emission of the ionised reflection were allowed to vary.
We assumed that all spectra are observed through a warm absorber with the same column density. 
 We also assume an identical power\--law index, normalisation of the distant reflection component, 
 inclination angle of the disk, and spin of the black hole. 
In addition, we fixed the radial dependence of the emission of the disk for the modelling of the highest flux state  (May
 2003) where it is basically unconstrained because of  the
 low contribution of the reflection  component to the total spectra.
From a statistical viewpoint, the model allows us to describe the data with $\chi^2=266.6$ for $d.o.f.=234$.
The observed spectra are compared with the model and the corresponding residuals  
 in Fig.~\ref{Fig4}. 
No systematic residuals remain around the iron line complex.

\section{Discussion}
\label{Dis}
Along the lines of our original observational goal, we have successfully pinpointed \object{PG~2112+059} 
 in its low state by acquiring a $\sim$25~ks short snapshot observation, and we performed a long 
 follow-up observation consisting of two $\sim$100~ks XMM-Newton observations within two and a half 
 weeks after the snapshot. 
Unfortunately, the two long observations were partly affected by  
 enhanced background radiation. 
Nevertheless, the number of collected events is significantly higher than in previous data, allowing a much 
 more detailed analysis of the source's low\--state spectral features.

As illustrated in Fig.~\ref{FigInt}, the main scientific challenge is the interpretation 
 of the excess emission or bump in the 2 to 5 keV energy range.
The request for consistency in the interpretation of the soft and hard parts 
 of the spectra constrains the issue.

\subsection{Distant cold reflector}
The spectra  of \object{PG~2112+059} taken in May 2007 exhibit a significant fluorescence 
line of neutral iron with a flux of 
${\rm F = \left( 6.4\pm2.8 \right) \times 10^{-7}\;\; photons\;cm^{-2}\;s^{-2}}$, which is in remarkable 
agreement with the flux found for a (2.6-$\rm \sigma$) line, 
 $F = \left( 5.0\pm3.2 \right)  \times 10^{-7} photons\;cm^{-2}\;s^{-2}$, 
 detected in the  November 2005 spectra  
(Schartel et al. \cite{Schartel2007}). 
The equivalent width found for the neutral iron line, ${\rm EW\;=\;110^{+60}_{-40}\;eV}$, is quite 
common for high luminosity, basically unabsorbed, type 1 objects.  
Jim\'{e}nez-Bail\'{o}n et al. (\cite{Jimenez2005}) detected fluorescence lines from neutral iron 
in  $\sim$50\% of the PG quasar sample, measure  ${\rm <EW>\;=\;80^{+30}_{-20}\;eV}$ with 
${\rm \sigma_{EW} < 40 \;eV}$. 
We interpret the line in the context of  a distant reflector in the sense that the reflecting material is far away 
 from the primary X-ray emitting source and the supermassive black hole. 
The distance is indeed so large that emission lines do not exhibit a measurable velocity 
 broadening and  the reflecting material does not become ionised. 
We accounted for the distance reflector following Nandra et al. (\cite{Nandra2007}).
The invariance of the line flux found in spectra of  \object{PG~2112+059} separated by 1.5 years 
 supports this interpretation.

\begin{figure}[th]
\centering
\includegraphics[width=6.5cm,angle=-90]{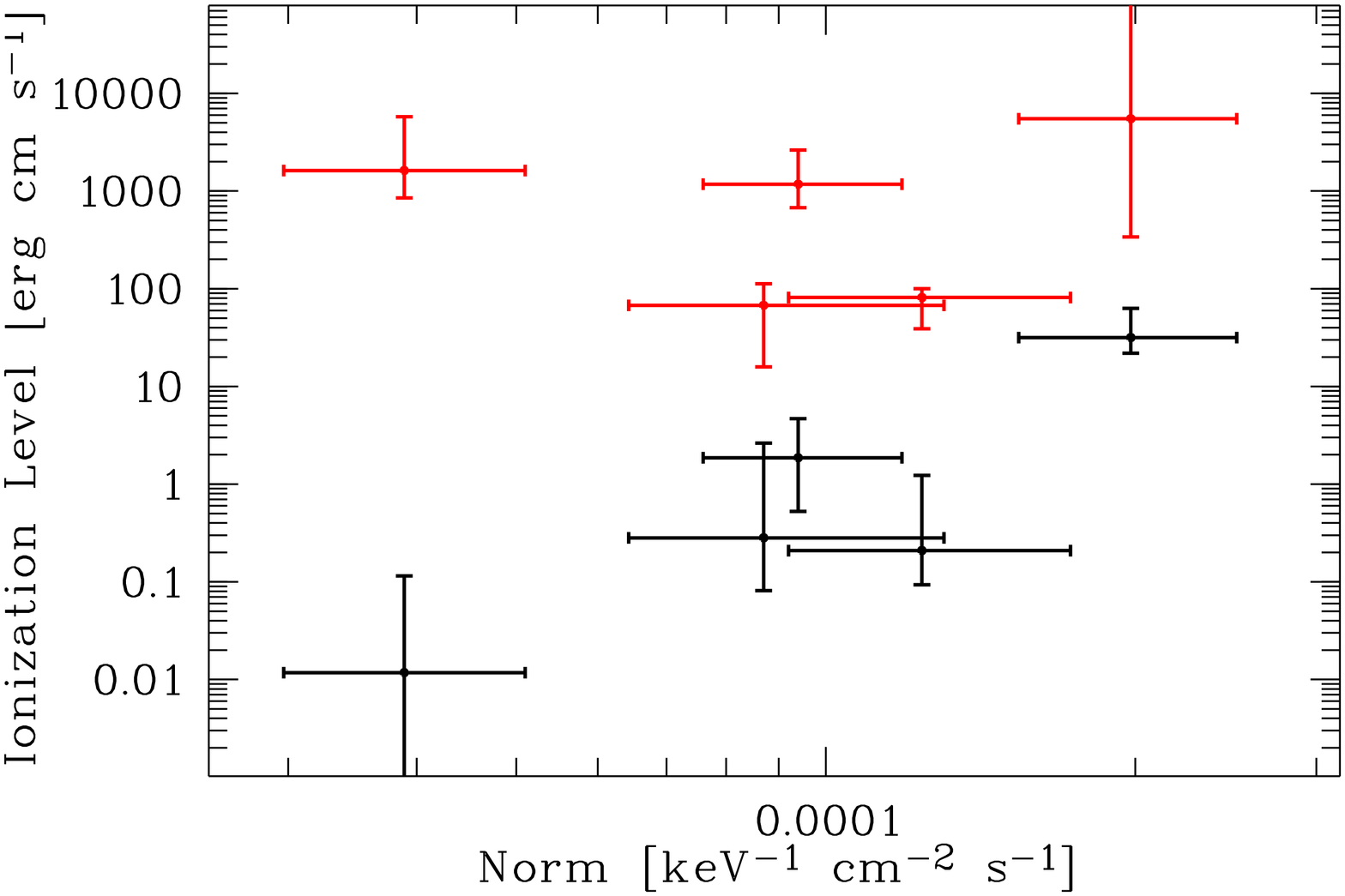}
\caption{
The ionisation parameter is plotted against the power\--law normalisation for the two 
 partially covering absorption layers that allow a joint description of the spectra 
 taken at different epochs and continuum flux levels (compare Table \ref{X_FITS_BLC}). 
As the power\--law indices were tied to each other, the normalisation is directly equivalent to the flux.
The highly ionised system is plotted in red and the low ionised system in black. }
 \label{ionab}
   \end{figure}

\subsection{Absorption}
The excess emission in the 2 to 5 keV energy range can be described by a power\--law source
 seen behind an absorber of high column density.
Models with neutral absorbers estimate the column densities to be greater than    
 $1.7 \times 10^{23}$ cm$^{-2}$ for the spectra of  \object{PG~2112+059} taken in 
 November 2005 (compare Table~\ref{FITS_IOC}). 
Ionised material must be of significantly higher column densities for the bump to be modelled  accurately.
Guainazzi,  Matt,  \& Perola (\cite{Guainazzi2005}) found that the column density  
 of obscured AGNs correlates well with the iron line equivalent width for   
 $N_H >10^{23}$ cm$^{-2}$ and flattens at lower densities to an almost constant value  
 of  $EW=113\pm13$eV, which is in agreement with the values found for low luminosity  
 unobscured quasars (Jim\'{e}nez-Bail\'{o}n et al. \cite{Jimenez2005}).
In the May 2007 spectra of \object{PG~2112+059}, we detected a neutral iron emission line 
 of $EW = 110^{+55}_{-40}$eV, which excludes models of the highest column densities considered. 

All attempts to describe the spectra taken at different epochs, by assuming  different flux 
 levels for the primary X-ray emitting source, require two layers of absorbing materials 
 of which at least one is ionised. 
The presented model (Table \ref{X_FITS_BLC}, fit 3) has two partially covering absorbing 
 systems \-- a low ionised system  
 with $N_H \sim 5 \times 10^{22} cm^{-2}$ and $\xi = \left( 0.01 -  30 \right) erg^{\;} s^{-1} cm $ and a highly ionised 
 system with $N_H \sim 3.5 \times 10^{23} cm^{-2}$ and $\xi =\left(  70 -  5500 \right) erg \, s^{-1} cm $.
Both systems can be described with a constant covering fraction of $cf \sim 0.8$.

The most similar absorbing system is that detected in \object{NGC 1365} (Risaliti et al. 
 \cite{Risaliti2005}, \cite{Risaliti2009A}, \cite{Risaliti2009B}).
A system of four iron absorption lines demonstrates the presence of a highly ionised absorber, of  column density $N_H \sim 5 \times 10^{23} cm^{-2}$ (Risaliti et al. \cite{Risaliti2005}).
In addition, Risaliti et al. (\cite{Risaliti2009B}) observed a variable (low ionised or 
 neutral) absorption system with $ N_H \sim 3.5 \times 10^{23} cm^{-2}$, which is probably a broad-line region cloud covering and uncovering the nucleus.
Risaliti et al (\cite{Risaliti2009A}) discussed two scenarios for explaining the 
 observed variability.
The one of special interest here consists of clouds with about the same size of 
 the X-ray emitting source and 
 assumes that, on average, few (1 $-$ 3) of these clouds transverse the line of sight. 
Depending on the fluctuations in the number and column density of the clouds, the 
 source exhibits a reflection-dominated or Compton-thin state.

The most striking difference between \object{NGC 1365} and \object{PG~2112+059} is 
 the amount of variability in the absorber, which is not detected at all for \object{PG~2112+059} as the data can 
 be described by a constant covering fraction. 
However, the supermassive black hole mass of \object{PG~2112+059} is 
 $\log M_{BH}/M_{solar}=9.0\pm0.1$ (Vestergaard \& Peterson \cite{Vestergaard2006}), 
 which is a factor $>$10 more massive than the supermassive black hole of 
 \object{NGC 1365} ($\log M_{BH}/M_{solar}= 7.2-7.8 \pm0.4$, Risaliti et al.\cite{Risaliti2009B}).
As the geometry scales with black hole mass, an absorption system analogue to the system of 
 \object{NGC 1365} could scale its size by a factor of 10.
If we assume a similar size for the clouds, we may expect basically a constant number of 
 clouds, $\sim 15$, along our line of sight, which would explain why we see no variability in the 
 covering factor.

In Fig.~\ref{ionab},  we show the ionisation parameter as a function of the power\--law 
 normalisation, which is proportional to the flux since the spectral slope is constant.
The ionisation parameters of the low ionised absorber are plotted in black.
The ionisation parameter is defined as $\xi = L n^{-1} r^{-2}$,  
 where $n$ is the electron density, $L$ is the luminosity  of the X-ray emitting source, 
 and $r$ is the distance.
Given the considerations above, we can assume that the absorbing clouds are at 
 a constant distance from the ionising source, which implies that
 $\xi \sim F n^{-1}$ 
 where F is the ionising flux.
For a constant electron density, we expect that $\xi \propto F$.
In Fig.~\ref{ionab} (black points), we find an increasing ionisation parameter 
 with increasing flux. 
However, a flux increase by a factor of 5 is accompanied by an ionisation 
 parameter increase of more than 250.
An increased flux causes a higher ionisation level and consequently  an increase in the electron density, which should even weaken {\bf the proportion between flux and ionisation parameter}.

The highly ionised absorber of \object{NGC 1365} (Risaliti et al. \cite{Risaliti2005}) 
 Is probably located close to the supermassive black hole, hence the high 
 value of its ionisation parameter. 
If this is the case, then the assumption of constant distance and density does 
 not hold, complicating the relationship between $\xi$ and $F$.
In Fig.~\ref{ionab}, we do not detect a correlation between 
 flux and ionisation parameter (red points).
It could also be that the ions responsible for absorption are already fully ionised 
 in the low flux state and therefore an increase in the flux would  not affect the ionisation parameter.

Although the observed properties can be explained by a scenario similar and scaled 
 according to the scenario developed by Risaliti et al. (\cite{Risaliti2009A}) 
 for \object{NGC 1365}, there are very important differences from \object{PG~2112+059}. 
Risaliti et al. find tight constraints on the variability within individual observations 
 and the absorption line identification is on a sound statistical basis. 
For \object{PG~2112+059}, the observations are separated by months and years, and there 
 is no clear measurement of the flux state between the different observations. 
In addition, the statistical data are insufficient for tracing the variability in individual absorption lines.

\begin{figure}[thb]
\centering
\includegraphics[width=6.5cm,angle=-90]{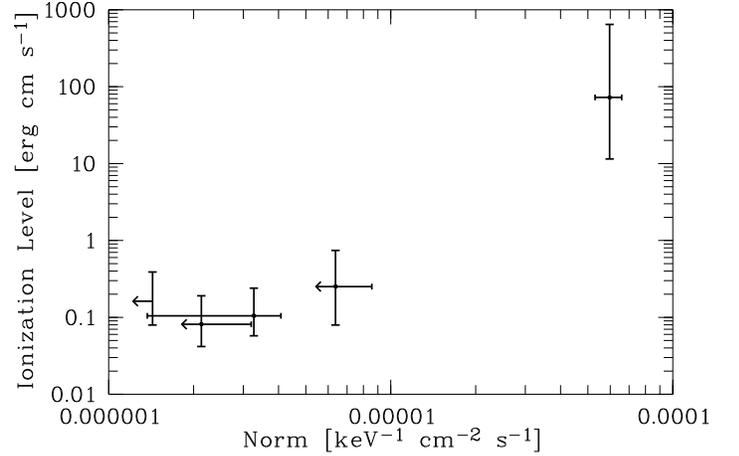}
\caption{
The ionisation parameter of the warm absorber is shown against the power\--law normalisation
 for the scenarios containing a  D\v{o}vciak, Karas  \& Yaqoob accretion disk for an attempt  to describe the spectra 
 taken at different epochs and flux levels  (compare Tables~\ref{X_FITS_BLC} and \ref{X_FITS_BL}).  
%Unfilled symbols represent upper limits to the flux estimate. 
As within each scenario, the power\--law indices were tied to each other, the normalisation 
 equals directly the flux. }
 \label{ionab3}
   \end{figure}

\subsection{Ionised reflection from an accretion disk}

Ionised reflection from an accretion disk allows us to describe the high energy bump of the spectra 
 taken of \object{PG~2112+059} in May 2007. 
Valid descriptions are obtained with the Laor (\cite{Laor1991}) blurring as well as 
 the blurring according to D\v{o}vciak, Karas  \& Yaqoob (\cite{Dovciak2004}).

We can jointly describe all five spectra of \object{PG~2112+059} by assuming ionised 
 reflection by an accretion disk blurred according to D\v{o}vciak, Karas \& Yaqoob 
 (\cite{Dovciak2004}), which  is observed behind a screen of ionised material and
 emission from a distant neutral reflector. 
We find an ionisation parameter of the warm absorber that follows the flux of the 
 continuum power laws (compare Fig.~\ref{ionab3}). 
Despite its large errors, the amplitude of the variation in the ionisation parameter  
 is equal to the amplitude of the variation in the flux.
With the exception of the high state, the index of the radial dependency of the disk 
 emission  agrees with the expected value of three. 
Hence, we consider this scenario to be physically meaningful and plausible.

\subsubsection{Spin of black hole}
\label{SoBH}
The data taken in May 2007 allow us to estimate the angular momentum of the black hole, 
 which is very high in all fits ($a/M>0.86$)
The data are consistent with a maximally rotating black hole.
Jointly fitting the five spectra taken at different epochs does not allow us
 to place tighter constraints on the black hole spin.

\subsubsection{Iron abundance}

The models used to describe the warm absorber, the distant reflector, and the 
 ionised reflection from the accretion disk include the iron abundance as a free parameter. 
We assumed that the iron abundance is the same in each of the spectral components.  
We note that the iron abundance of the 
 absorption scenarios is within the range of abundances found for scenarios assuming 
 ionised reflection by the disk when the blurring is modelled according to
  D\v{o}vciak, Karas \& Yaqoob (\cite{Dovciak2004}),
whereas the scenarios that 
 treat blurring according to Laor (\cite{Laor1991}) show significantly
  higher abundance values.
Restricting ourselves to the joint modelling of the spectra taken at different 
 epochs, we find $Fe/Fe_{\odot} = 1.5 - 2.2$ for the absorption scenarios,  
 $Fe/Fe_{\odot} = 1.6 - 1.9 $ for the ionised reflection blurred according to 
 D\v{o}vciak, Karas \& Yaqoob (\cite{Dovciak2004}), 
 and $Fe/Fe_{\odot} = 2.9 - 3.2$ if blurred according to Laor (\cite{Laor1991}).  
These super-solar abundances are frequently measured in both  low-z and high-z AGNs. 
However, the high metallicity value may be an artefact as different physical models in terms of  limb brightening were applied.

\subsection{Accretion rate}

Witt, Czerny \&  {\.Z}ycki (\cite{Witt1997}) discussed an accretion disk with a 
 hot continuous corona. 
The authors assumed that the corona itself accretes and is therefore powered by 
 the release of the gravitational energy and cooled by radiative interaction with the disk. 
In this model, the radial infall is accompanied by a strong vertical outflow. 
In the context of our observations, the most important prediction is that at a given 
 radius the corona forms only for accretion rates higher than a limiting value and 
 that the fraction of the energy dissipated in the corona decreases with increasing accretion rate.

Considering  optical and especially ultraviolet data we clearly trace a decrease 
 in the flux from 2005 to 2007, which for the UVW2 filter is of the order of 10\%. 
The decrease of the U band flux is of the order of 8\%, i.e., comparable with the 
 variation found for the UV. 
Unfortunately, we do not have an ultraviolet observation that was performed during the high state in 2003. 
However, based on the estimated optical flux and assuming a constant ratio of optical 
 to ultraviolet flux and accounting for our full error budget, we can assume for 2003 an 
 UV flux not more that 20\% higher than in 2005. 
Assuming that the ultraviolet emission is proportional to the accretion rate, we would 
 expect a similar variability in the X-ray flux. 
However, the observed X-ray variability is larger than expected and, furthermore, 
 in 2005, when the UV emission was 10\% higher than in 2007, the X-ray count 
 rate was about 40\% lower than in May 2007.
Therefore, we cannot explain the X-ray variability with changes in the accretion rate
  using the model of Witt, Czerny \&  {\.Z}ycki (\cite{Witt1997}).

\subsection{Light bending model}
Our findings for \object{PG~2112+059} can be understood in the context of the light bending model for the 
 spectral properties of accreting black holes developed by Miniutti \& Fabian (\cite{Miniutti2004}). 
In this model  primary emission is generated by  a ring-like source that emits hard X-ray radiation  isotropically 
 with constant luminosity in the form of a power law. 
The source is centred on a maximally rotating Kerr black hole rotation axis at a height of between only 1$r_g$ 
 and 20$r_g$ above the  equatorial plane. 
In this model, changes in the observed flux and changes in the ratio of the  power law continuum to
 reflection from the disk are explained by variations in the height of the primary source: 
 as the primary source lowers its height above the black hole, more light will be gravitationally 
 bended towards the disk lowering the observed at infinity continuum flux, while  the flux in the reflection increases.

\begin{figure}[h]
\centering
\includegraphics[width=6.5cm,angle=-90]{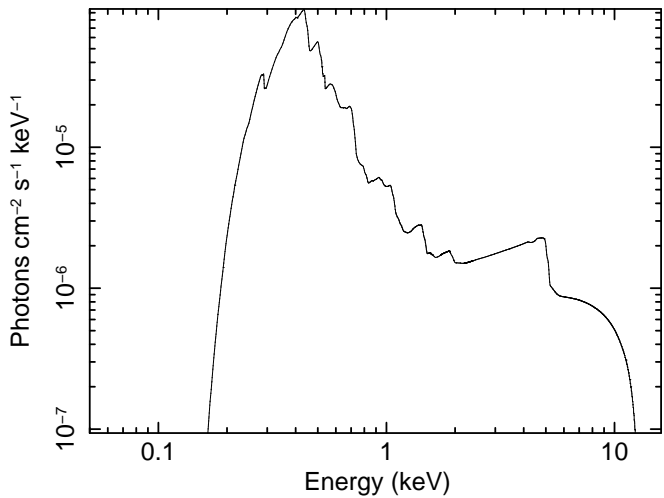}
\caption{
The figure shows best\--fit models assuming ionised reflection by the accretion disk at three epochs. 
The ionised absorber and the distant reflection components are not shown for clarity. 
The upper-panel shows the model for the May 2003 spectra, when the source showed the highest flux. 
The middle panel shows the model for the May 2007 spectra, when both components significantly contributed to the total emission.
The lower panel shows the model for the May 2005, when \object{PG~2112+059} was in a deep minimum state.  }
 \label{Model}
   \end{figure}

The light bending model assumes a maximally spinning black hole. Modelling the spectra of 
 \object{PG~2112+059} with the black hole spin as a free parameter we obtain very high estimates 
 for the spin in agreement with the assumption of a maximally spinning black hole.
The main arguments in favour of the Miniutti \& Fabian light bending model are the observed low 
 states in 2005 and 2007. 
Blurring the disk reflection spectra according to Laor, the primary power\--law 
 components disappear completely, and we can even speak of ``reflection-dominated''  
 (Table~\ref{X_FITS_BLC} and Table~\ref{X_FITS_BL} fit 4 and Fig.~\ref{Model}).  
These spectra correspond to the regime 1 in the light bending model, where the primary source is 
 located at only 2 to 4 $r_s$ from the black hole:  the spectra are dominated by the reflection component 
 and the continuum is strongly suppressed.  
The model of Laor blurring assumes a maximal rotating black hole. 
We obtain an index for the power 
 law radial dependence of the disk emission significantly higher than 3, 
which supports our
 interpretation as exactly this is predicted
\begin{center}
\includegraphics[width=6.5cm,angle=-90]{12389f10.ps}
\end{center}
\begin{center}
\includegraphics[width=6.5cm,angle=-90]{12389f11.ps}
\end{center}
 by the light bending model.
The most accurately estimated disk inclination angle is in agreement with 60$^o$, and therefore we can directly 
 compare the measured  flux ratios with Fig. 2 (lower panel)  in  Miniutti \& Fabian (\cite{Miniutti2004}).  
If we assume that the observed low states correspond to phase 1 of the model, then the observed continuum 
 flux for the high state is in disagreement with the model assumptions.

Thus, light bending is the simplest, physically consistent model that can explain our observations of 
 \object{PG~2112+059}.

\section{Conclusions}
\label{Con}

We have analysed observations that demonstrate that XMM-Newton can identify low states using a  short snapshot observation and long follow-up observations.
 An absorption scenario, similar to the scenario proposed by Risaliti et al. (\cite{Risaliti2009A}) 
 for \object{NGC 1365} allows us to understand the X-ray spectra of \object{PG~2112+059} taken 
 at different flux states.
The system consists of two layers of ionised material, both covering the source by 80\%.
They show column densities of  $N_H \sim 5 \times 10^{22} cm^{-2}$ and  
 $N_H \sim 3.5 \times 10^{23} cm^{-2}$. 
The first is moderately ionised and its ionisation levels follow the flux changes. 
The latter is highly ionised and shows not correlation with the flux of the source.

A scenario with reflection by an ionised disk in the context of a light bending model 
 (Miniutti \& Fabian \cite{Miniutti2004}) also provides a convincing physical interpretation
 of the different source states. 
The warm absorber ionisation is consistent with being correlated with the flux of the source, 
 which provides an additional degree of self-consistency to the overall reflection-based model.

A joint analysis of spectra taken at different epochs (i.e., corresponding to different emission states) provide  results that are useful to constraining the emission scenarios but do not allow us to break the degeneracy  
 between  complex absorption scenarios and reflection by an ionised disk in the case of  \object{PG~2112+059}.
Future high resolution observations of \object{PG~2112+059} carried out by 
 next-generation X-ray telescopes  will be crucial to the detailed study of the absorption, 
 the continuum shape, and the strength of the reflection components.

\begin{acknowledgements}
The work is based on observations obtained with XMM-Newton, an ESA science mission with  
 instruments and contributions directly funded by ESA Member States and NASA.
We thank K. Nandra for providing  the pexmon model.
Lucia Ballo acknowledges support from the Spanish Ministry of Science and Innovation through 
 a "Juan de la Cierva" fellowship.  
Financial support for this work was provided by the Spanish Ministry of Science and Innovation, 
 through research grant ESP2006-13608-C02-01.
Enrico Piconcelli acknowledges support under ASI/INAF contract I/088/06/0.
We thank the anonymous referee for many fruitful comments and suggestions.
\end{acknowledgements}

\begin{appendix}
\label{A}

\section{Exposure details}
\label{A1}

In Table~\ref{TOBS}, the exposure details for the XMM-Newton observations 
 of \object{PG~2112+059} in 2007 are given. 
For each exposure identifier, we report the instrument, filter, start time, 
and exposure time.

\begin{table}[ht]
\caption[]{\label{TOBS}
XMM-Newton 2007 observations of \object{PG~2112+059}}
\begin{tabular}{rllcc}
\hline \\
N.$^{(1)}$ & I.$^{(2)}$ & Filter & Start         & Duration \\
          &        &        & Day\&Time     &             \\
                 &        &        &  [UT]   & [ks] \\ \\
\hline \hline \\ 
\multicolumn{5}{c}{\underline{May 2007 / 5005006}} \\ 
1  & M1 & thin 1          &  3 at 21:45:41 &  24.1 \\
2  & M2 & thick           &  3 at 21:45:41 &  24.1 \\
3  & pn & thin 1          &  3 at 22:08:23 &  22.1 \\
6  & OM & V               &  3 at 21:50:19 &   2.2 \\
7  & OM & U-NoBar         &  3 at 22:32:06 &   2.2 \\
8  & OM & B               &  3 at 23:13:53 &   2.2 \\
9  & OM & V GRISM 2       &  4 at 23:55:40 &   4.0 \\
10 & OM & UVW1            &  4 at 01:07:27 &   2.2 \\
11 & OM & UVM2            &  4 at 01:49:14 &   2.2 \\
12 & OM & UVW2            &  4 at 02:31:01 &   2.8 \\
13 & OM & UV GRISM 1      &  4 at 03:22:48 &   3.8 \\ 
\multicolumn{5}{c}{\underline{May 2007 / 5005007}} \\ 
1  & M1 & thin 1          & 19 at 10:03:47 & 100.0  \\
2  & M2 & thick           & 19 at 10:03:47 & 100.0  \\
3  & pn & thin 1          & 19 at 10:26:29 &  98.4  \\
6  & OM & V               & 19 at 10:08:25 & 2 $\times$ 3.3 \\
9  & OM & U-NoBar         & 19 at 12:08:39 & 2 $\times$ 3.3 \\
12 & OM & B               & 19 at 14:08:53 & 2 $\times$ 3.3 \\
15 & OM & V GRISM 2       & 19 at 16:09:07 & 4 $\times$ 5.0 \\
19 & OM & UVW1            & 19 at 22:48:45 & 2 $\times$ 3.3 \\
20 & OM & UVW1            & 20 at 00:48:59 &            3.5 \\
21 & OM & UVM2            & 20 at 01:52:26 & 3 $\times$ 3.9 \\
25 & OM & UVW2            & 20 at 05:23:59 & 3 $\times$ 3.5 \\
28 & OM & UV GRISM 1      & 20 at 08:34:20 & 2 $\times$ 5.0 \\ 
30 & OM & UV GRISM 1      & 20 at 08:34:20 & 2 $\times$ 4.7 \\ 
\multicolumn{5}{c}{\underline{May 2007 / 5005008}} \\ 
1  & M1 & thin 1          & 21 at 11:19:23 &  91.9  \\
2  & M2 & thick           & 21 at 11:40:27 &  90.6  \\
3  & pn & thin 1          & 21 at 11:42:58 &  90.2  \\
6  & OM & V               & 21 at 10:00:35 & 2 $\times$ 3.3 \\
9  & OM & U-NoBar         & 21 at 12:00:49 & 2 $\times$ 3.3 \\
12 & OM & B               & 21 at 14:01:03 & 2 $\times$ 3.3 \\
15 & OM & V GRISM 2       & 21 at 16:01:17 & 4 $\times$ 5.0 \\
19 & OM & UVW1            & 21 at 22:40:43 & 2 $\times$ 3.3 \\
21 & OM & UVW1            & 21 at 00:40:57 &            3.5 \\
22 & OM & UVM2            & 22 at 01:44:24 & 3 $\times$ 3.7 \\
25 & OM & UVW2            & 22 at 05:07:39 & 2 $\times$ 3.5 \\
29 & OM & UV GRISM 1      & 22 at 08:18:00 &            4.7 \\ 
30 & OM & UV GRISM 1      & 22 at 08:18:00 &            5.0 \\ 
\multicolumn{5}{c}{\underline{November 2007 / 5005009}} \\ 
1  & M1 & thin 1         &  5 at 09:52:51 &  91.9  \\
2  & M2 & thick          &  5 at 09:52:51 &  90.6  \\
3  & pn & thin 1         &  5 at 10:15:12 &  90.2  \\
6  & OM & V              &  5 at 09:57:29 &            2.0 \\
9  & OM & U-NoBar        &  5 at 10:35:56 &            2.5 \\
12 & OM & B              &  5 at 11:22:43 &            2.0 \\
15 & OM & V GRISM 2      &  5 at 12:01:10 & 2 $\times$ 4.0 \\
19 & OM & UVW1           &  5 at 14:24:44 & 2 $\times$ 3.5 \\
23 & OM & UVM2           &  5 at 16:31:38 & 2 $\times$ 3.5 \\
26 & OM & UVW2           &  5 at 18:38:32 & 2 $\times$ 3.5 \\
29 & OM & UV GRISM 1     &  5 at 20:45:26 &            4.0 \\ 
30 & OM & UV GRISM 1     &  5 at 21:57:13 &            4.4 \\ 
31 & OM & UV GRISM 1     &  5 at 23:16:17 &            3.4 \\ \\
\hline \\
\end{tabular}

(1): Exposure identifier: 
     If OM performed more than one exposure with the same specifications 
      (filter, grism and exposure time) then only exposure identifier and 
      start time of the first exposure is provided.

(2): instrument, where M1 stands for MOS 1 and M2 for MOS2.
\end{table}

\section{Processing details}
\label{A2}

\begin{table}
\caption[]{\label{TSCR} 
Screening for low background level, source position and background extraction area }
\begin{tabular}{llrr}
\hline \\ 
\multicolumn{1}{l}{N.$^{(1)}$} & \multicolumn{1}{l}{EPIC$^{(2)}$}  & \multicolumn{2}{c}{Net Source} \\
 &  & Count Rate$^{(3)}$  & Exposure Time$^{(4)}$  \\  
 &  & [$\rm 10^{-2} s^{-1}$] & [$\rm ks$] \\  \\
\hline \hline \\ 
\multicolumn{4}{c}{Observation from 14.5.2003 / OI$^{(5)}$: 0150610201} \\ \\
1 & pn & 13.12$\pm$0.44 &   7.1 \\ 
2 & M1 &  3.92$\pm$0.21 &  8.9  \\
3 & M2 &  4.12$\pm$0.22 &  9.2  \\ \\
\multicolumn{4}{c}{Observation from 20.11.2005 / OI$^{(5)}$: 0300310201} \\ \\
1 & M1 &  0.39$\pm$0.03 &  72.4 \\
2 & M2 &  0.37$\pm$0.03 &  73.7 \\
3 & pn &  1.26$\pm$0.06 &  63.7 \\ \\
\multicolumn{4}{c}{Observation from 03.05.2007 / OI$^{(5)}$: 0500500601} \\ \\
1 & M1 &  0.65$\pm$0.06 &  22.9 \\
2 & M2 &  0.54$\pm$0.06 &  23.7 \\
3 & pn &  1.90$\pm$0.13 &  13.8 \\ \\
\multicolumn{4}{c}{Observation from 19.05.2007 / OI$^{(5)}$: 0500500701} \\ \\
1 & M1 &  0.55$\pm$0.04 &  54.8 \\
2 & M2 &  0.54$\pm$0.03 &  57.8 \\
3 & pn &  1.91$\pm$0.08 &  40.4 \\ \\
\multicolumn{4}{c}{Observation from 21.05.2007 / OI$^{(5)}$: 0500500801} \\ \\
1 & M1 &  0.68$\pm$0.03 &  79.4 \\
2 & M2 &  0.54$\pm$0.03 &  89.5 \\
3 & pn &  2.08$\pm$0.07 &  62.8 \\ \\
\multicolumn{4}{c}{Observation from 05.11.2007 / OI$^{(5)}$: 0500500901} \\ \\
1 & M1 &  0.43$\pm$0.03 &  50.6 \\
2 & M2 &  0.45$\pm$0.03 &  50.8 \\
3 & pn &  1.51$\pm$0.07 &  42.5 \\ \\
\\  \hline \\   
\end{tabular}  

(1): exposure identifier;
(2): M1 stands MOS1 and M2 for MOS2;
(3): background corrected source count rate in the energy range of 
        0.2-12.0 keV for pn and 0.2-10.0 keV for MOS, respectively;
(4): accumulated exposure time.
(5): observation identifier
\end{table}

The EPIC and OM data were processesd as described in Schartel et al. (\cite{Schartel2007}). 
The EPIC data were analysed to determine the time ranges with low radiation background level 
 according to Piconcelli et al. (\cite{Piconcelli2005}). 
Special care was taken to ensure that all data of each observation were screened 
 and that the source and background counts were extracted in the same way 
 such that artificial differences were avoided. 

The source counts were extracted from a circular region of radius $ r=20^{''}$ 
 centred on the source positions determined by eye 
 for each observation and exposure. 
The eye-determined positions agree with the optical position 
of \object{PG~2112+059}, given the pointing accuracy of XMM-Newton of $\sim 1.5^{''}$.

The screening for low background intervals was based on an annulus centred 
 on the source position. 
For both cameras, we chose an inner radius of $r_{inner}=1^{'}$.
Reflecting the different detector geometry, we chose an outer radius of 
 $r_{outer}=14^{'}$ for the MOS cameras and of $r_{outer}=11^{'}$ for the pn camera.
The events were binned with 100s. 
We considered an energy range from 0.2 to 10 keV for the two MOS cameras and 
 from 0.2 to 12 keV for the pn camera.
Good time intervals were characterised by a count rate $CR < 2.5 s^{-1}$ for MOS 
 and $CR < 5.0 s^{-1}$ for pn with the exception of the May 2003 pn observation, 
 where we applied $CR < 10.0 s^{-1}$ reflecting the high and soft background conditions.

The background spectra for the MOS cameras were collected from an annulus around the 
 source position with an inner radius of $r_{inner}=40^{''}$ 
 and an outer radius of $r_{outer}=105^{''}$. 
A circular region with a radius of $r = 30^{''}$ was used for the background 
 determination for the pn spectra. 
The region was centred on about $\rm RA=21h14m49s$ and $\rm Dec=6d7m22s$ for 
 all observations with the exception of the November 2005 observation, where the centre was at
$\rm RA=21h14m57s$ and $\rm Dec=6d8m4s$ reflecting the position angle of the observation.

In Table~\ref{TSCR} give the accumulated low background time for each exposure and 
 the corresponding net source count rate.

\end{appendix}

\listofobjects

\end{document}